\documentclass{aa}  
\usepackage{graphicx}
  \usepackage[fleqn]{amsmath}
\usepackage{txfonts}
\usepackage{natbib}
\bibpunct{(}{)}{;}{a}{}{,}	
\begin{document}
   \title{Gravitational fragmentation caught in the act: the filamentary Musca molecular cloud\thanks{This publication is based on data acquired with the Atacama Pathfinder Experiment (APEX). APEX is a collaboration between the Max-Planck-Institut f\"ur Radioastronomie, the European Southern Observatory, and the Onsala Space Observatory (Max-Planck programme ID M-085.F-0027).}}
   \author{J. Kainulainen	\inst{1},
           A. Hacar 		\inst{2},
           J. Alves			\inst{2}, 
           H. Beuther		\inst{1},
           H. Bouy			\inst{3}, \and
	  M. Tafalla		\inst{4}
          }


   \institute{Max-Planck-Institute for Astronomy, K\"onigstuhl 17, 69117 Heidelberg, Germany \\
              \email{jtkainul@mpia.de}         
         \and 
              Institute for Astronomy, University of Vienna, T\"urkenschanzstrasse 17, 1180 Vienna, Austria
         \and
              Centro de Astrobiolog\'ia, INTA-CSIC, PO Box 78, 28691 Villanueva de la Ca\~nada, Madrid, Spain 
        	\and
             Observatorio Astron\'omico Nacional (IGN), Alfonso XII 3, 28014 Madrid, Spain 
                          }
   \date{Received ; accepted }
  \abstract
   {Filamentary structures are common in molecular clouds. Explaining how they fragment to dense cores is a missing step in understanding their role in star formation.}
   {We perform a case study of whether low-mass filaments are close-to hydrostatic prior to their fragmentation, and whether their fragmentation agrees with gravitational fragmentation models. For this, we study the $\sim$6.5 pc long Musca molecular cloud that is an ideal candidate for a filament at an early stage of fragmentation.}
   {We employ dust extinction mapping in conjunction with near-infrared $JHK_\mathrm{S}$ band data from the CTIO/NEWFIRM instrument, and 870 $\mu$m dust continuum emission data from the APEX/LABOCA instrument, to estimate column densities in Musca. We use the data to identify fragments from the cloud and to determine the radial density distribution of its filamentary part. We compare the cloud's morphology with $^{13}$CO and C$^{18}$O line emission observed with the APEX/SHeFI instrument.}
   {The Musca cloud is pronouncedly fragmented at its ends, but harbours a remarkably well-defined, $\sim$1.6 pc long filament in its Center region. The line mass of the filament is 21-31 M$_\odot$ pc$^{-1}$ and $FWHM$ 0.07 pc. The radial profile of the filament can be fitted with a Plummer profile that has the power-index of $2.6 \pm 11$\%, flatter than that of an infinite hydrostatic filament. The profile can also be fitted with a hydrostatic cylinder truncated by external pressure. These models imply a central density of $\sim$5-10$\times 10^4$ cm$^{-3}$. The fragments in the cloud have a mean separation of $\sim$0.4 pc, in agreement with gravitational fragmentation. These properties, together with the subsonic and velocity-coherent nature of the cloud, suggest a scenario in which an initially hydrostatic cloud is currently gravitationally fragmenting. The fragmentation has started a few tenths of a Myr ago from the ends of the cloud, leaving its centre yet relatively non-fragmented, possibly because of gravitational focusing in a finite geometry.}
{}
  \authorrunning{J. Kainulainen et al.}
  \maketitle


\section{Introduction} 
\label{sec:intro}


Filamentary structures have been known to be common morphological features of the interstellar medium (ISM) for decades \citep[e.g.,][]{bar27, sch79, dob05, and14}.  Observational works examining the physical properties of filaments have built a picture in which filamentary clouds fragment into denser clumps and/or dense cores that can further collapse to form new stars \citep[e.g.,][]{bal87, miz95, mye09, and10, and14, jac10, sch10, kai13}. This phenomenology is apparent across a wide range of size- and mass-scales, spanning $\sim$0.1-100 pc in size and $\sim$1-10$^5$ M$_\odot$ in mass \citep[e.g.,][]{pin10, hac11, sch10, hil11, hac13, goo14, rag14}. The new, sensitive dust emission observations by the \emph{Herschel Space Observatory} have also manifested the universality of filamentary structures in molecular clouds, reinforcing a hypothesis that they may provide a dominant pathway to star formation in molecular clouds \citep[for a review, see][]{and14}. Explaining how  filamentary structures evolve and form stars is then a necessary step in understanding the exact role of this pathway.


Analytical studies have established a viable framework for the existence of filamentary structures, regardless of their origin, as hydrostatic, self-gravitating, thermally pressurised cylinders \citep{ost64, fis12a}. These cylinders can fragment in time-scales of $\sim$Myr when perturbed \citep{inu92, inu97, fis12a}, and even small density perturbations in the filament can collapse faster than the global collapse takes place in it \citep{pon11}. This gravitational fragmentation provides one viable model to explain the "dense cores" observed in filaments. The details of the fragmentation process in this model depend most crucially on the line mass of the filament at its initial condition, i.e., at the equilibrium configuration. The line mass at equilibrium, and hence the fragmentation process, can be affected by the possible pressure confinement by the surrounding environment \citep{fis12a}, support provided by magnetic fields \citep{fie00a, fie00b}, accretion of gas onto the filament \citep{hei13a, hei13b}, temperature profile \citep{rec13}, and rotation \citep{rec14}. For a more comprehensive review, we refer to \citet{and14}.


One key open question related to filaments is whether \emph{i) }the filaments are close to hydrostatic objects prior to their fragmentation, in which case they first have to detach from the turbulent regime of the cloud and then fragment, or \emph{ii) }the evolutionary picture of filaments is dynamic and hydrostatic conditions are never prevalent in the cloud. This question can be addressed by first carefully identifying filaments that are at the earliest stages of their evolution, possibly not yet fragmented, and by comparing their structure with the predictions of the hydrostatic equilibrium models. However in practice, doing this is severely hampered by the fact that \emph{non-fragmented filaments that carry the imprints of the initial condition are rare}. This alone indicates that the phase in which a filament is non-fragmented and yet CO-emitting (so that it would be identified as a molecular cloud) is relatively short. While several recent works have made efforts to estimate the structure of filaments in nearby clouds \citep[e.g.,][]{arz11, fis12b, juv12b, pal13}, none of these works to our knowledge has identified and analysed specifically non-fragmented filaments that would allow an assessment of the filament structure close to its initial condition.

Another key question is how the filaments fragment into dense cores that form stars. Analytic models predict the initially hydrostatic, infinitely long filament to fragment through gravitational instability at a wavelength of roughly four times the filament scale-height \citep{inu92}, in a time-scale comparable to the sound-crossing time \citep[order of a Myr, e.g.,][]{fis12a}. For finite-sized filaments, it has been hypothezised that the local collapse time-scale varies as a function of the position along the filament, and that the local collapse proceeds most rapidly at the ends of the filament \citep{bas83, pon11}. While testing these predictions is an active goal of observational studies \citep[e.g.,][]{jac10, schm10, bus13, kai13, tak13}, it is of great interest to find and characterise the fragmentation in \emph{as simple as possible} filaments to isolate different processes playing a role.  

The main technical difficulty in addressing the questions above is the general difficulty of tracing the distribution of molecular gas in the interstellar medium: all commonly-used tracers (dust emission, extinction, molecular line emission) have their specific problems and none of them can alone trace the wide range of column densities present in filaments. An additional difficulty is introduced by the small spatial scale of the filaments and their fragments. Filaments typically show widths that are on the order of 0.1 pc \citep[e.g.,][]{bar27, arz11}; fragments are similar in size. Consequently, single-dish observations can only study the structure of filaments in adequate detail in the most nearby molecular clouds (distances $\lesssim 200$ pc). As a result, it remains unknown what the physical structure of filaments is like \emph{prior to their fragmentation} and how exactly their fragmentation proceeds from that initial state to star formation.

\begin{figure*}
   \centering
\includegraphics[bb = 20 50 650 622, clip=true, width=0.53\textwidth]{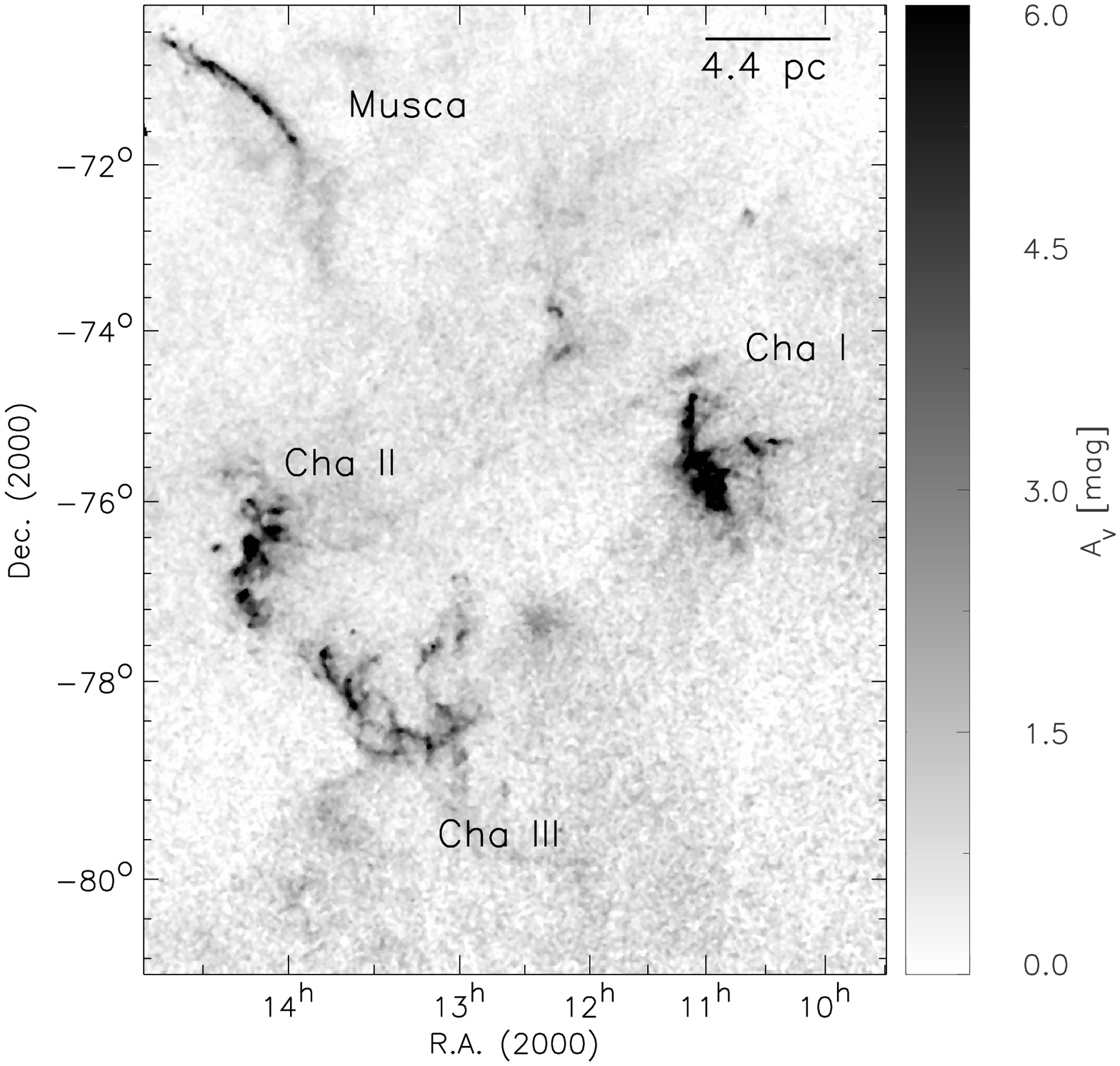}\includegraphics[bb = 60 50 635 640, clip=true, width=0.47\textwidth]{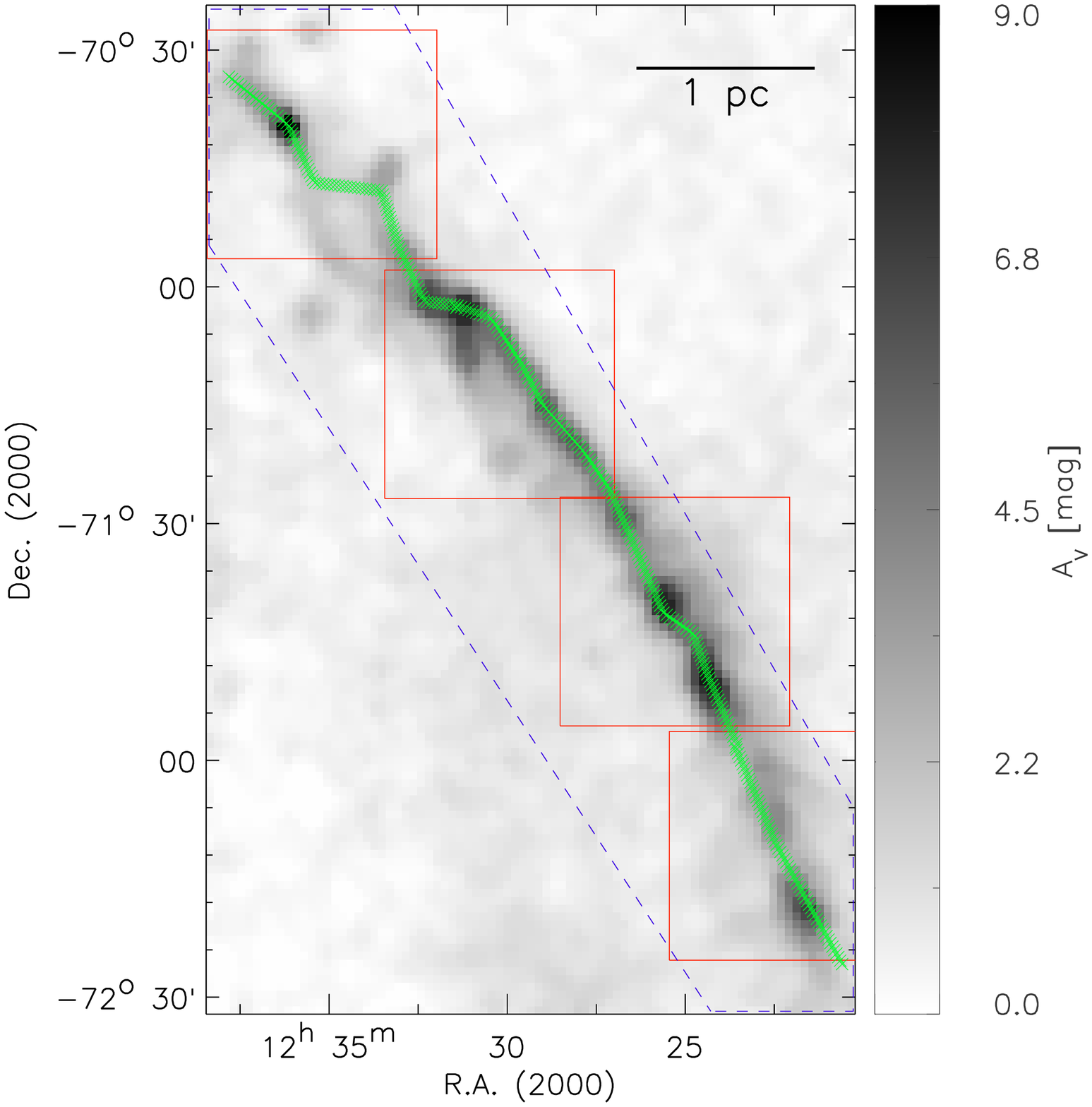} 
      \caption{\emph{Left: }Dust extinction map of the Chamaeleon-Musca region, based on 2MASS data \citep{kai09}. \emph{Right: }Zoom-in to the Musca cloud, showing the instrument footprints of the observations employed in this work: NEWFIRM/NIR data (red rectangles), LABOCA/870 $\mu$m continuum emission (blue dashed lines, approximate coverage), SHeFI/C$^{18}$O line emission \citep[green crosses, from][]{hac15}.
              }
         \label{fig:musca_large}
   \end{figure*}

To make progress in characterising the early evolution and fragmentation of filaments, we target in this study the remarkably long ($\sim$6.5 pc), highly rectilinear \object{Musca} molecular cloud. The cloud is located in the Musca-Chamaeleon cloud complex (see Fig. \ref{fig:musca_large}) at the distance of $\sim$150 pc \citep{knu98}. The cloud has been identified to harbor only a very small number of cold cores given its spatial extent \citep[][]{vil94, juv10, juv11, juv12b}, and it contains only one T Tauri star candidate, attesting to its low current star formation activity \citep{vil94}. We have recently shown using CO line observations that Musca is entirely subsonic and velocity-coherent \citep{hac15}. Contrasting this apparent quiescence, we have also shown that the probability distribution of column densities in the cloud shows a power-law-like excess of high-column densities, a feature common for star-forming molecular clouds \citep{kai09, kai11}. Similarly, the cloud's volume density distribution is more top-heavy than that of other nearby starless clouds, suggesting that Musca may be close to active star formation \citep{kai14}, or possibly in transition between non-star-forming and star-forming states.

The above properties make Musca a perfect candidate to be in the earliest phases of fragmentation. In this paper, we present a detailed study of the column density structure of Musca. We use archival near-infrared (NIR) observations and 870 $\mu$m dust emission observations to map column densities through the cloud and to characterise its internal structure. We then place the results in the context of hydrostatic equilibrium models, establish their suitability to describe the cloud, study its fragmentation, and finally propose an evolutionary scenario for the cloud.


\section{Data and methods}           
\label{sec:data}

\subsection{NIR data and dust extinction mapping}
\label{subsec:nirdata}


The Musca cloud was observed with the NIR wide-field camera NEWFIRM \citep{2003SPIE.4841..525A} mounted on the Blanco 4-m telescope at the Cerro Tololo Inter-American Observatory between 2011 April 11 and 13. We retrieved the raw data and associated calibration files from the NOAO public archive. Six pointings (5 covering the entire cloud and 1 off the cloud) were obtained in the $J$, $H$ and $K_\mathrm{s}$ broad-band filters. Each pointing was observed in dither mode. Table~\ref{tab:newfirm} gives a summary of the observation strategy. The instrument footprints are shown in Fig. \ref{fig:musca_large}, right panel; note that only four frames fall along the main part of Musca shown in the figure.

The individual raw images were processed using an updated version of \emph{Alambic} \citep{2002SPIE.4847..123V}, 
a software suite developped and optimized for the processing of large multi-CCD imagers. \emph{Alambic} includes standard processing procedures such as overscan and bias subtraction, flat-field correction, bad pixel 
masking, CCD-to-CCD gain harmonization, sky-subtraction, and de-stripping. Aperture and PSF photometry were extracted from the individual images using SExtractor \citep{1996A&AS..117..393B} and PSFEx \citep{PSFEx}. The individual catalogs were then astrometrically registered and aligned on the same photometric scale using Scamp \citep{2006ASPC..351..112B}, and the final mosaic produced using SWarp \citep{2002ASPC..281..228B}. The photometric zero-points of the merged catalog were derived using the 2MASS catalog. The final absolute astrometric accuracy is expected to be better than 0\farcs1 and the absolute photometric accuracy (rms of the individual images zeropoints) of the order of 5\%. The NEWFIRM data were complemented with $J$, $H$, and $K_\mathrm{s}$ band photometric data from the 2MASS archive \citep[][]{skr06} to cover a wider field towards Musca. The area covered by the NEWFIRM frames cover the area used in all analyses of this paper; the 2MASS data is merely used to complete the outskirts of a wide-field extinction map (Fig. \ref{fig:musca_large}, right panel).

\begin{table*}
\centering
\caption{Overview of NEWFIRM observations \label{tab:newfirm}}
\begin{tabular}{lccc}\hline\hline
Band      & Number of dithered positions & Number of coadds & Single coadd exp. time \\
J  & 20 & 2 & 30~s \\ 
H  & 25 & 16 & 4~s \\
Ks & 30 & 12 & 5~s \\
\hline
\end{tabular}
\end{table*}



We used the $JHK_\mathrm{S}$ band data in conjunction with the \textsf{nicest} color-excess mapping technique \citep{lom09} to estimate the dust extinction through the Musca cloud. For the details of the technique we refer to \citet{lom09}. The technique is based on comparing NIR colors ($H-K$ and $J-H$) and surface density of the stars that shine through the cloud to those from a field that is assumed to be free from extinction. We note that since the photometric depths of the NEWFIRM and 2MASS data are different, these parameters were determined separately for the two data sets.  
In implementing \textsf{nicest}, a relative extinction law between $JHK_\mathrm{s}$ bands has to be assumed. We chose the empirical extinction law of \citet{car89}
\begin{equation}
\tau_\mathrm{K} = 0.60 \times \tau_\mathrm{H} = 0.40 \times \tau_\mathrm{J}.
\end{equation}
We emphasize that the technique estimates the dust extinction in NIR. We convert the estimates of NIR extinction to visual extinction by adopting the ratio between optical depths in NIR and visual wavelengths \citep{car89}
\begin{equation}
\tau_\mathrm{V} = 8.77 \times \tau_\mathrm{K}.
\end{equation}
%
%
The visual extinctions are finally used to estimate the total hydrogen (atomic + molecular) column densities using the ratio measured by \citet{sav77} and \citet{boh78} \citep[see also][]{vuo03, gue09}
\begin{equation}
N_\mathrm{H} = 2N(\mathrm{H}_2)+N(\mathrm{H}) = 1.9 \times 10^{21} \mathrm{\ cm}^{-2}\ \times A_\mathrm{V} \ \mathrm{mag}^{-1}   
\label{eq:bohlin}
\end{equation}
The final extinction map of Musca is presented in Fig. \ref{fig:nir-av}, with higher-detail blow-ups shown in Fig. \ref{fig:nir-av-bu}. The map has the resolution of 45 arcsec and uncertainty at low column densities of $\sigma$($A_\mathrm{V}$)=0.2 mag. We analyse the column and volume density structure of the cloud using this map in Sections \ref{subsec:cd_structure} and \ref{subsec:physical_models}.

\subsection{LABOCA dust emission data}

In addition to extinction maps, we carried out in May 2010 870 $\mu$m continuum observations of the Musca cloud using the LABOCA bolometric camera \citep{SIR09} at the APEX telescope. The observations consisted of fourteen individual raster-spiral maps combined to obtain a large-scale mosaic of approximately 2.2 by 0.4 deg$^2$ covering the total extension of this cloud. Guided by our previous extinction maps of Musca \citep{kai09}, the location, coverage, and overlap of these submaps were selected to optimise the sensitivity of the final mosaic along the main axis of the cloud. Skydips plus flux calibrations and pointing corrections in nearby sources were carried out every 1.5-2 hours. Using the BoA software \citep{SCH12}, the data reduction process followed the standard method described by \citet{BEL11} in order to recover the extended emission in the cloud. Convolved to a final resolution of 30 arcsec, the resulting map presents a typical rms of 17 mJy beam$^{-1}$ and is shown in Fig.\ref{fig:nir-av}.

\subsection{SHeFI $^{13}$CO and C$^{18}$O line emission data}
\label{subsec:shefi}

We employ in the discussion of this paper $^{13}$CO ($J=2-1$) and C$^{18}$O ($J=2-1$) line emission data of the Musca cloud from \citet{hac15}, to which we refer for the description of the observations and data reduction. The data were gathered with the SHeFI instrument at the ESO/APEX telescope in Chile. The spatial resolution of the data is $28\farcs5$ and the velocity resolution 0.10 km s$^{-1}$ or 0.17 km s$^{-1}$ (the receiver backend was changed during the observations). The spectra typically have a main beam temperature rms of 0.14 K. Line emission was observed in 305 positions along a curve that approximately traces the crest of the Musca cloud (see Fig. \ref{fig:musca_large}). The data were used to determine the central velocities, $V_\mathrm{LSR}$, and velocity dispersions, $\sigma _v$, along the filament. They were also used to estimate the non-thermal contributions to the observed line-widths, $\sigma _\mathrm{NT}$ and $\sigma _\mathrm{NT, corr}$ (opacity-corrected), assuming a temperature of 10 K. For the details of the derivation of these values we refer to \citet{hac15}.

\section{Results}           
\label{sec:results}

\subsection{Column density structure of the Musca filament}
\label{subsec:cd_structure}

\begin{figure*}
   \centering
\includegraphics[bb = 95 80 780 800, clip=true, width=\textwidth]{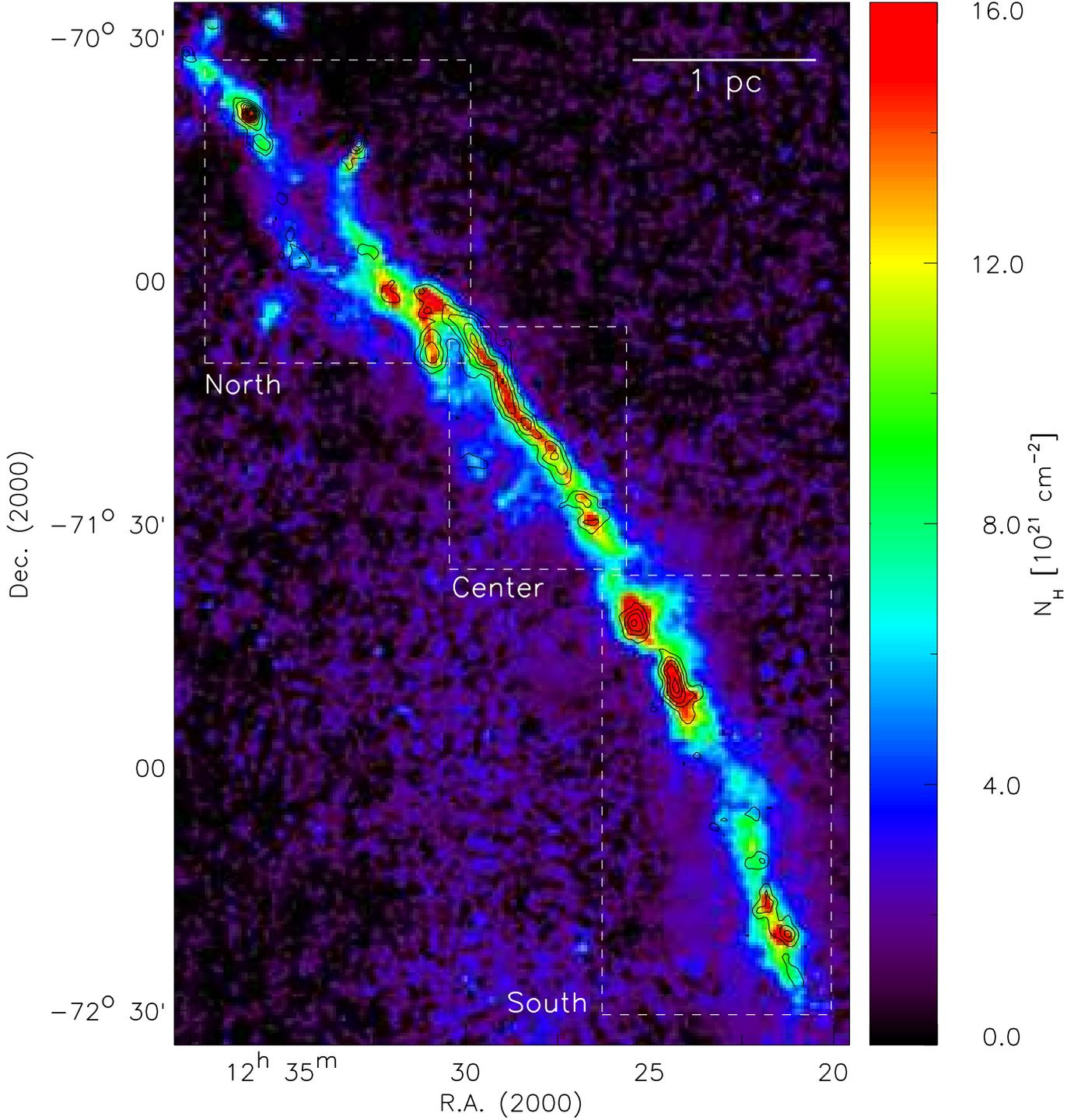} 
      \caption{Column density map of Musca (colour scale), derived using near-infrared dust extinction mapping and data from NEWFIRM and 2MASS. The spatial resolution of the map is $FWHM = 45\arcsec$ and the sensitivity $\sigma (N_\mathrm{H}) \approx 0.4 \times 10^{21}$ cm$^{-2}$. The white boxes indicate three regions that are shown in blow-ups in Fig. \ref{fig:nir-av-bu}. The contours show the 870 $\mu$m continuum emission contours, observed with the APEX/LABOCA instrument and smoothed down to the resolution of the extinction data. The contours are drawn at the intervals of 0.03 mJy/beam, which is the 3-$\sigma$ rms noise of the smoothed data. 
              }
         \label{fig:nir-av}
   \end{figure*}
\begin{figure*}
   \centering
\includegraphics[bb = 50 20 1360 650, clip=true, width=\textwidth]{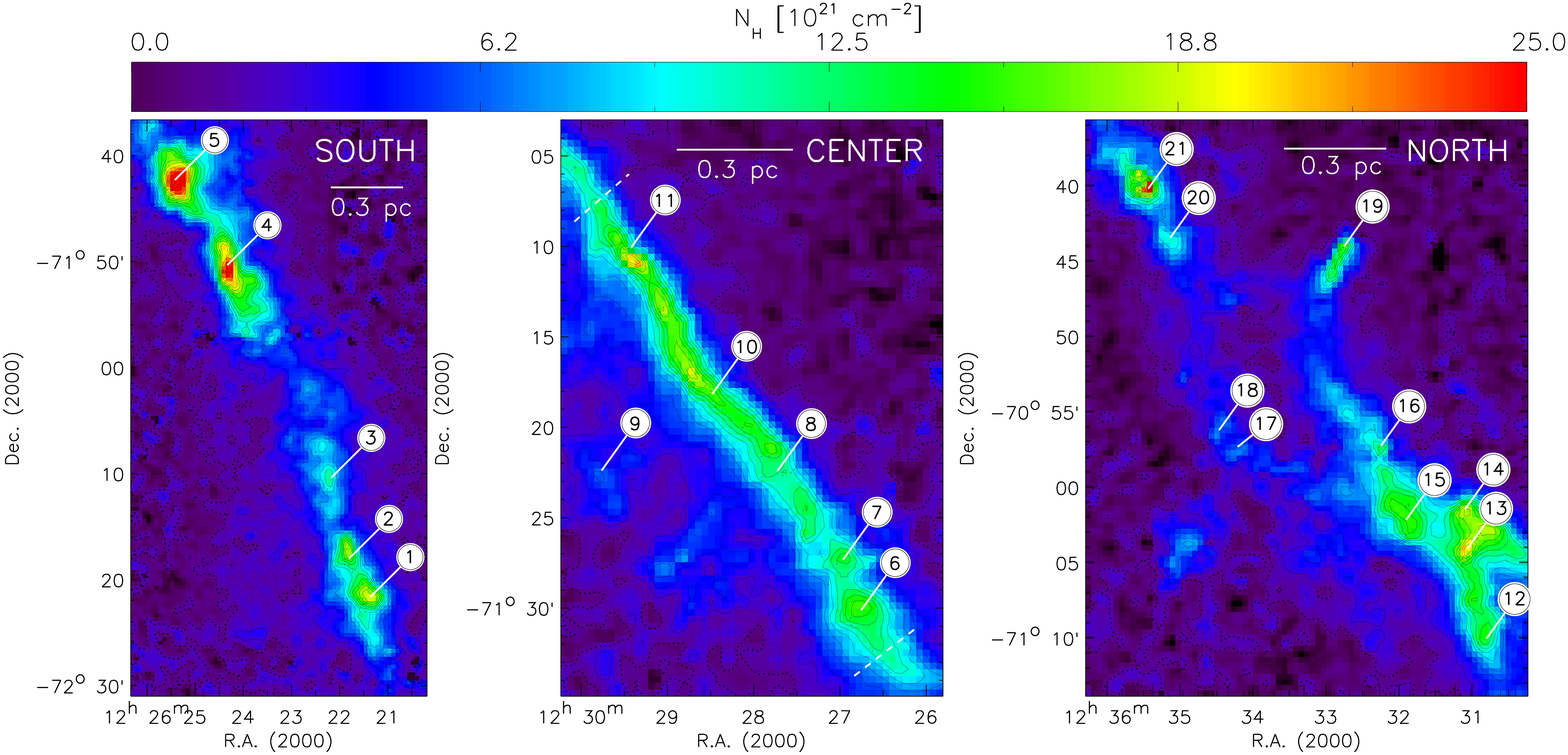}  
      \caption{Blow-ups from the NIR-derived column density map of the Musca cloud (Fig. \ref{fig:nir-av}). The contours start at $N_\mathrm{H} = 1.5 \times 10^{21}$ cm$^{-2}$ and the step is $1.5 \times 10^{21}$ cm$^{-2}$. The lowest contour is drawn with a dotted line. The dashed lines in the Center region indicate the 1.6 pc velocity-coherent filament whose radial structure is modelled in Section \ref{subsec:physical_models}. The numbered signs refer to the fragments identified within the cloud (see Section \ref{subsec:cd_structure}).
              }
         \label{fig:nir-av-bu}
   \end{figure*}

Figure \ref{fig:nir-av} shows the column density map of the Musca cloud derived from the NIR dust extinction data. Figure \ref{fig:nir-av-bu} shows zoom-ins of the map. The highest column densities in the map are about $35 \times 10^{21}$ cm$^{-2}$. The mean column density above $2 \times 10^{21}$ cm$^{-2}$ is $\langle N_\mathrm{H} \rangle = 4.5 \times 10^{21}$ cm$^{-2}$. Figure \ref{fig:nir-av} also shows the contours of the LABOCA 870 $\mu$m emission data. The emission contours show good morphological correspondence with the NIR-derived column density map, evidencing that no column density peaks remain undetected from the NIR-derived data, e.g., because of the saturation of extinction or resolution effects. In general, the relationship between the extinction- and emission-derived column densities shows a high scatter, likely due to the spatial filtering of the dust emission data and their lower signal-to-noise compared to extinction data; the relationship is presented in Appendix \ref{app:laboca-vs-nir}.

We estimated the total mass of the cloud from the NIR-derived column densities by summing up all column densities inside the contours\footnote{To describe how the chosen contour level affects the results, we give the masses and line masses corresponding to three different contour levels.} of $N_\mathrm{H} = \{1.5, 3, 4.5\} \times 10^{21}$ cm$^{-2}$ (see Fig. \ref{fig:nir-av-bu} for the contours). This resulted in the total masses of $M = \{281, 174, 132\}$ M$_\odot$. We used the total masses to calculate equivalent\footnote{We use the term ``equivalent line mass'', because the cloud is partly fragmented (North and South regions) and does not resemble a well-defined filament throughout its length. In this case, ``line mass'' alone would be a somewhat misleading term.} line masses, $M^\mathrm{e}_\mathrm{l}$, of the entire Musca cloud. The length of Musca for all thresholds is about $l = 6.5$ pc, resulting in $M^\mathrm{e}_\mathrm{l} = \{43, 27, 20\}$ M$_\odot$ pc$^{-1}$. The inclination angle, $i$, of the cloud is not known, and for reference, we adopt here $i = 0$. The inclination affects the line masses through a factor of $\cos{i}$, and hence, the values we give represent upper limits. Note that in addition to the inclination uncertainty, the uncertainty of the line mass is significantly affected by distance uncertainty, which is at least some $\sim$10\%. 

The column density map (Fig. \ref{fig:nir-av}) reveals numerous details within the large-scale Musca cloud. We divided the cloud into three regions, namely South, Center, and North, based on their morphology. The detailed column density maps of these regions are shown in Fig. \ref{fig:nir-av-bu}.

%
%

The column density map (and emission data) show that Musca contains several small-scale condensations. To have a simple, quantitative census of these structures, we identify ``fragments'' in the observed region. We emphasize that our purpose here is merely to have a first look at the small-scale substructures in the cloud; a simple fragment definition will suffice for this purpose. With this in mind, we first smooth the resolution of the emission data down to the resolution of the extinction data (45\arcsec) and identify fragment candidates as local emission maxima. In identifying the local maxima, we use contour steps of 0.03 mJy. If the fragment candidate is smaller than $45\arcsec$ in diameter, it is rejected. As a second step, the fragment candidates are inspected against the local maxima of the extinction map. Local maxima are determined from the extinction data using contour intervals of 1.5$\times 10^{21}$ cm$^{-2}$. If the lowest contour that outlines a fragment candidate does not overlap with the lowest contour outlining an extinction maximum, the candidate is rejected. The pixel with maximum emission value is recorded as the location of the fragment. As a result of these steps, our fragments are essentially local emission maxima that have counterparts in the extinction map. The procedure results in identifying 21 fragments from the observed area.

We estimated the masses associated with the fragments by summing up all column densities from the NIR-derived data within circular apertures. The radii of the apertures, listed in Table \ref{tab:cores}, were chosen by eye. The fragment masses span 0.3-7.1 M$_\odot$ (Table \ref{tab:cores}). We note that all column densities within the apertures were included in the calculation, i.e., possible "background" was not considered; it is not our goal to analyse the internal structure of the fragments or their relation to their surroundings. Finally, we roughly estimated the mean densities of the fragments from their area and mass, assuming spherical symmetry (Table \ref{tab:cores}).


%
%

In the following, we describe the specific properties of the South, Center, and North regions separately. The South region is clearly fragmented into five core-like structures, the most massive of which contain several solar masses. We measured the separation between the fragments as distances between neighbouring fragments; in case of the South (and Center) this is trivial, because the fragments are arranged into a line along the filament crest. The mean separation in South is $0.78\pm 0.6$ pc. Note that the mean is strongly affected by the long distance between fragments \#3 and 4; without the separation between those two fragments the mean peak separation is $0.50\pm 0.2$ pc. The total mass of the South region above $N_\mathrm{H} = \{1.5, 3, 4.5\} \times 10^{21}$ cm$^{-2}$ are $M = \{96, 88, 74\}$ M$_\odot$. The region is about 2.7 pc long, resulting in the equivalent line masses of $M^\mathrm{e}_\mathrm{l} = \{36, 33, 27\}$ M$_\odot$ pc$^{-1}$.  Finally, we note that fragments \#4 and \#5 practically compose the object studied as ``a filament'' by \citet[][]{juv12b}. It appears to us that this region is clearly fragmented, and therefore, the axisymmetric cuts across the region measure in effect a combination of the profile of the fragments and that of the collapsing filament, not the profile of the filament itself. 


The Center region of the Musca cloud consists of a 1.6 pc long filament that shows less sub-structure than the South region and has a relatively constant peak column density of $N_\mathrm{H} \approx 10-15 \times 10^{21}$ cm$^{-2}$ along its crest. The line mass of this filament using the column density thresholds of $N_\mathrm{H} = \{1.5, 3, 4.5\} \times 10^{21}$ cm$^{-2}$ results in $M_\mathrm{l} = \{31, 26, 21\}$ M$_\odot$ pc$^{-1}$. Thus, its line mass is practically the same as that of the entire Musca cloud on average and that of the South and North (see below) regions. The $FWHM$ of the filament is 0.07 pc, measured from its mean column density profile (see Section \ref{subsec:physical_models}). There are five fragments along the main filament and one about 0.5 pc off the filament. The fragments along the filament have very low contrast to their surroundings compared to those in the South (see the quantification of the column density contrast later in this Section). The mean separation of the five fragments along the crest is $0.35 \pm 0.13$ pc (fragment \#9 was not included in the analysis). We consider physical models for the Center filament in Section \ref{subsec:physical_models}.


The North region of the Musca cloud is clearly fragmented, similarly to the South region. In the south-west corner of the region, there is a complex area containing four fragments. This region is characterised by two velocity components, likely resulting from two overlapping structures \citep[][]{hac15}. In the middle of the North region, there is a curious structure that consists of two thin ($N_\mathrm{H} \approx 3 \times 10^{21}$ cm$^{-2}$) structures parallel to each other (a bubble-like structure), and of two peaks (fragments \#19 in northwest and fragments \#17 and 18 in southeast) that are elongated almost in perpendicular direction to those two structures. Unfortunately, we did not cover this region with molecular line data, and thus the nature of the structure remains unclear. There is a strong local maximum at the north-east corner of the region. This peak coincides with the location of the only candidate young star \citep[T Tauri star candidate,][]{vil94} in the cloud. The North region harbours ten fragments (Table \ref{tab:cores}). These fragments are not aligned along a line as clearly as fragments in Center and South. Acknowledging this caveat, we calculated the separations along two routes that connect the fragments (in the order of the fragment number): one including fragment \#19 and excluding fragments \#17 and 18, and the other excluding fragment \#19 and including fragments \#17 and 18. This resulted in mean separations of $0.24 \pm 0.16$ pc and $0.25 \pm 0.17$ pc, respectively. The total masses of the region above $N_\mathrm{H} = \{1.5, 3, 4.5\} \times 10^{21}$ cm$^{-2}$ are $M = \{59, 47, 37\}$ M$_\odot$. With the approximate length of 2.0 pc of the region, these result in the equivalent line masses of $M^\mathrm{e}_\mathrm{l} = \{30, 24, 19\}$ M$_\odot$ pc$^{-1}$. 

\begin{figure}
   \centering
\includegraphics[width=\columnwidth]{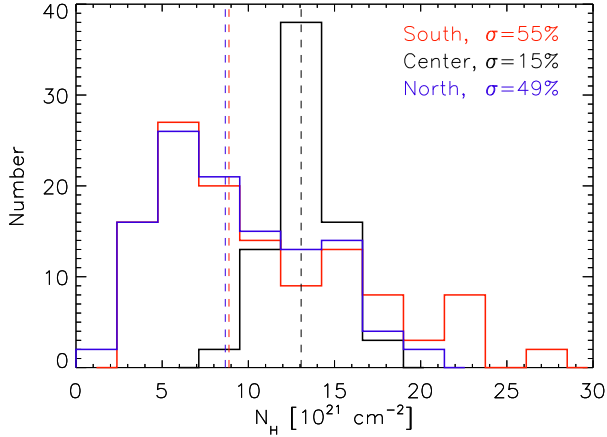} 
      \caption{Frequency distribution of column density values along the approximate crest of the Musca cloud for South (red), Center (black), and North (blue) regions. The dashed lines indicate the mean column densities. The relative standard deviation of the column density values in each region is shown in the panel.
              }
         \label{fig:av_variance}
   \end{figure}

We emphasize that the North and South regions contain high-contrast density enhancements compared to the smoother, filamentary Center region; even though we identify "fragments" from the Center region, their column density contrast against their surrounding, more elongated gas component is very low. We further quantify this morphological difference by measuring the standard deviation of column density (extinction) values along the approximate crest of the cloud (see Fig. \ref{fig:musca_large} right panel). The histograms of the extinction values along the crest in different regions are shown in Fig. \ref{fig:av_variance}. The figure shows that the distributions of North and South regions are clearly different from that of the Center region: the North and South peak at around $N_\mathrm{H} \approx 6 \times 10^{21}$ cm$^{-2}$ and have strongly skewed distributions, while the Center peaks at around $N_\mathrm{H} \approx 13 \times 10^{21}$ cm$^{-2}$ and has a narrow distribution. The relative standard deviations of the extinction values along the cloud crest are: 55\% for South, 49\% for North, and 15\% for the Center. This shows that the column density variation in the Center is clearly smaller, by a factor of $\sim$3, than in the North and South.


\begin{table*}
\caption{Physical properties of fragments in Musca}             
\label{tab:cores}     
\centering                    
\begin{tabular}{l c c c c c c c c}     
\hline\hline            
\# 	& R.A. 	& Dec. 	& $N_\mathrm{H}^\mathrm{peak}$  & $F_\mathrm{870, peak}$ & $R$	&	 $M_\mathrm{NIR} $ 	 & $\langle n \rangle$  & Region  \\   
	&		&		&	[$10^{21}$ cm$^{-2}$]			& [Jy beam$^{-1}$]		& [pc]	&	[M$_\odot$]	&	[$10^{4}$ cm$^{-3}$] &	\\
\hline            
   1 	&   12:21:26	& -72:21:06 	& 23 		& 0.15	& 0.07		& 2.5		& 4		& South    \\   
   2 	&   12:21:54	& -72:17:02 	& 19 		& 0.11	& 0.07 		& 1.7		& 3		& South 	\\   
   3\tablefootmark{a} 	&   12:22:19	& -72:10:24 	&   17 		& 0.10	& 0.06 		& 1.0		& 2		& South 	\\
   4 	&   12:24:32	& -71:50:23 	& 27 		& 0.22	& 0.08 		& 3.5		& 3		& South 	\\
   5 	&   12:25:37 	& -71:42:14  	& 34 		& 0.18	& 0.11 		& 7.1		& 3		& South 	\\    
\hline
   6 	&   12:26:46	& -71:30:00 	& 15 		& 0.10	& 0.07 		& 1.7		& 3		& Center 	\\
   7 	&   12:26:59 	& -71:27:41  	& 17 		& 0.10	& 0.06 		& 1.2		& 3		& Center 	\\    
   8\tablefootmark{b} 	&   12:27:45 	& -71:22:07  	& 15 		& 0.10	& 0.05 		& 1.3		& 3		& Center 	\\    
   9\tablefootmark{c} 	&   12:29:46 	& -71:22:42  	& 6 		& 0.05	& 0.04 		& 0.5		& 1		& Center 	\\        
   10\tablefootmark{a} 	&   12:28:30 	& -71:18:13  	& 17 		& 0.12	& 0.04 		& 1.4		& 4		& Center 	\\
   11\tablefootmark{a} 	&   12:29:26 	& -71:10:41  	& 20 		& 0.17	& 0.05 		& 1.4		& 3		& Center 	\\    
\hline
   12 	&   12:30:48	& -71:10:08 	& 15 		& 0.14	& 0.05 		& 0.9		& 4		& North 	\\
   13 	&   12:31:03 	& -71:04:32  	& 19 		& 0.10	& 0.04 		& 1.1		& 6		& North 	\\    
   14 	&   12:31:05	& -71:01:54 	& 19 		& 0.09	& 0.04 		& 0.9		& 5		& North 	\\
   15 	&   12:31:53	& -71:02:08 	& 17 		& 0.07	& 0.07 		& 1.8		& 3		& North 	\\
   16 	&   12:32:13 	& -70:57:49  	& 13 		& 0.06	& 0.04 		& 0.6		& 4		& North 	\\    
   17 	&   12:34:09 	& -70:57:40  	& 7 		& 0.06	& 0.04 		& 0.3		& 2		& North 	\\    
   18 	&   12:34:24	& -70:56:11 	& 8 		& 0.08	& 0.04 		& 0.3		& 2		& North 	\\
   19 	&   12:32:41 	& -70:44:21  	& 15 		& 0.19	& 0.06 		& 1.0		& 2		& North 	\\    
   20 	&   12:35:00 	& -70:43:54  	& 11 		& 0.12	& 0.05 		& 0.6		& 3		& North 	\\    
   21 	&   12:35:17 	& -70:40:10  	& 30 		& 0.62	& 0.05 		& 1.1		& 5		& North 	\\    
\hline               
\end{tabular}
\tablefoot{
\tablefoottext{a}{The peak value of dust emission from the fragment is more than 45\arcsec offset from the associated extinction peak. The structure is considered as a fragment, because the contours outlining the local maximum in emission overlap with the contours outlining a local maximum in extinction. The mass estimate is based on extinction data; the aperture for the mass calculation was centered on the nearest extinction maximum.}
\tablefoottext{b}{The contours defining the local maximum in emission map overlap with two local extinction maxima. The mass estimate is based on extinction data; the aperture for the mass calculation was centered on the nearest extinction maximum.}
\tablefoottext{c}{The fragment is offset by about 0.5 pc from the filament crest. Consequently, we do not include the fragment into the fragmentation analysis.}
}
\end{table*}

\subsection{Physical models for the \emph{Center} filament}
\label{subsec:physical_models}


The column density maps of the Center region (Figs. \ref{fig:nir-av}, \ref{fig:nir-av-bu}) revealed an about 1.6 pc long, relatively non-fragmented filament. In this section, we examine the radial structure of this filament and consider physical models for it.


We estimated the mean column density profile of the Center filament from radial cuts across it. For this purpose, we first defined the filament crest by connecting the local maxima within the $N_\mathrm{H} = 8 \times 10^{21}$ cm$^{-2}$ contour with lines. The filament crest is shown in Fig. \ref{app:crest} and the points defining it are given in Table \ref{tab:crest}. We then sampled the radial profile with cuts perpendicular to the crest. The positions of the cuts along the crest were separated by 4 pixels to ensure that the profiles at the crest position are not correlated. Consequently, the $1.6$ pc long filament was sampled with 20 profiles. 


Figure \ref{fig:radialmodels} shows the resulting mean profiles, given separately for the west and east sides of the filament. Importantly, the profiles show that the filament is not symmetric, but the west side of it is steeper than the east side. The CO spectra show two velocity components on the east side of the filament crest \citep[cf.,][]{hac15}, suggesting that another, faint structure overlaps with the main crest at the east side. This asymmetry is also seen at larger scales in the extinction map of the region (see Fig. \ref{fig:musca_large}). Therefore, we consider that the west side profile reflects the true filament profile more correctly. It is noteworthy that even if Musca seems to represent one of the simplest filamentary clouds we know, its radial profile is asymmetric with respect to the main axis when resolved in high detail. Similarly, as shown in Section \ref{subsec:cd_structure}, small scale variations are present even in the simple Musca filament. These phenomena are likely present in more complex filamentary clouds, such as L1495-B213 \citep[e.g.,][]{schm10} and IC5146 \citep[e.g.,][]{arz11}. This advices caution when analysing profiles of complex filaments by averaging individual profiles and need to be addressed for the correct physical interpretation of the derived parameters. In the following, we quantify the radial structure of the Center filament in Musca by fitting different physically motivated models to its observed profile. 

\subsubsection{Isolated, infinite filament in hydrostatic equilibrium}
\label{subsec:radial_structure_simple}

We first modelled the 1.6 pc long filament in the Center region in the context of an infinite, isolated cylinder in equilibrium between thermal and gravitational pressures \citep{ost64}. Such an equilibrium configuration has the line mass
\begin{equation}
M_\mathrm{l, crit} = \frac{2kT}{\mu m_\mathrm{H}G} = \frac{2c_\mathrm{s}^2}{G} \approx 16 \mathrm{\ M}_\odot \mathrm{\ pc}^{-1} \times (\frac{T}{10 \mathrm{\ K}}), 
\label{eq:ostriker}
\end{equation}
where $c_\mathrm{s}$ is the sound speed. This indeed is close to the observed line mass of the Center filament ($M_\mathrm{l} = 21$-31 M$_\odot$). We note that it is possible that non-thermal motions provide additional support for a filament, making its critical line mass higher. An estimate of this ``turbulent support'' can be calculated from Eq. \ref{eq:ostriker} by replacing the sound speed with an effective sound speed, $c_\mathrm{s, eff}$, that results from the total line width in the cloud. The non-thermal line-width of C$^{18}$O) is about 0.7$c_\mathrm{s}$, yielding $c_\mathrm{s, eff} = \sqrt{1+0.7^2} c_\mathrm{s} = 1.22 c_\mathrm{s}$. The critical line mass is affected by the square of this, yielding 24 M$_\odot$ pc$^{-1}$. 

The volume density profile of the infinite, hydrostatic filament in equilibrium is \citep{ost64}
\begin{equation}
\rho (r) = \frac{ \rho_\mathrm{c}    }{  (1+(r/(\sqrt{8}r_\mathrm{0}))^2)^{p / 2}   },
\end{equation}
where $r_\mathrm{0}$ is the scale radius as defined by \citet{ost64}, $\rho_\mathrm{c}$ the central density, and the exponent $p$=4. Note that if the exponent is \emph{not} 4, the radial profile does \emph{not} describe an equilibrium solution. Using the generic, Plummer-like profile is however useful, because it allows for quantification of $r_\mathrm{0}$, $\rho_\mathrm{c}$, and $p$ through a fit to an observed profile, and from therein, examination of the validity of the model. The scale radius is coupled to the central density and temperature, $T$, through
\begin{equation}
\rho_\mathrm{c} = \frac{kT}{\mu m_\mathrm{H}4\pi G r_\mathrm{0}^2}.
\label{eq:arz_profile_ncenter_theory}
\end{equation}
The column density profile corresponding to the volume density profile is \citep[e.g.,][]{arz11}
\begin{equation}
N(r) =  A_\mathrm{p} \frac{\rho_\mathrm{c} \sqrt{8} r_\mathrm{0}}{\big(1 + \big(  r / (\sqrt{8} r_\mathrm{0})   \big)^2   \big)  ^{(p - 1) / 2}}.
\label{eq:arz}
\end{equation}
In the equation, $A_\mathrm{p} = \frac{1}{\cos{i}} \int_{-\infty}^{\infty} \frac{du}{(1+u^2)^{p/2}}$, with $i$ as the inclination angle of the filament, which we assume to be zero. 

\begin{figure*}
   \centering
\includegraphics[width=\columnwidth]{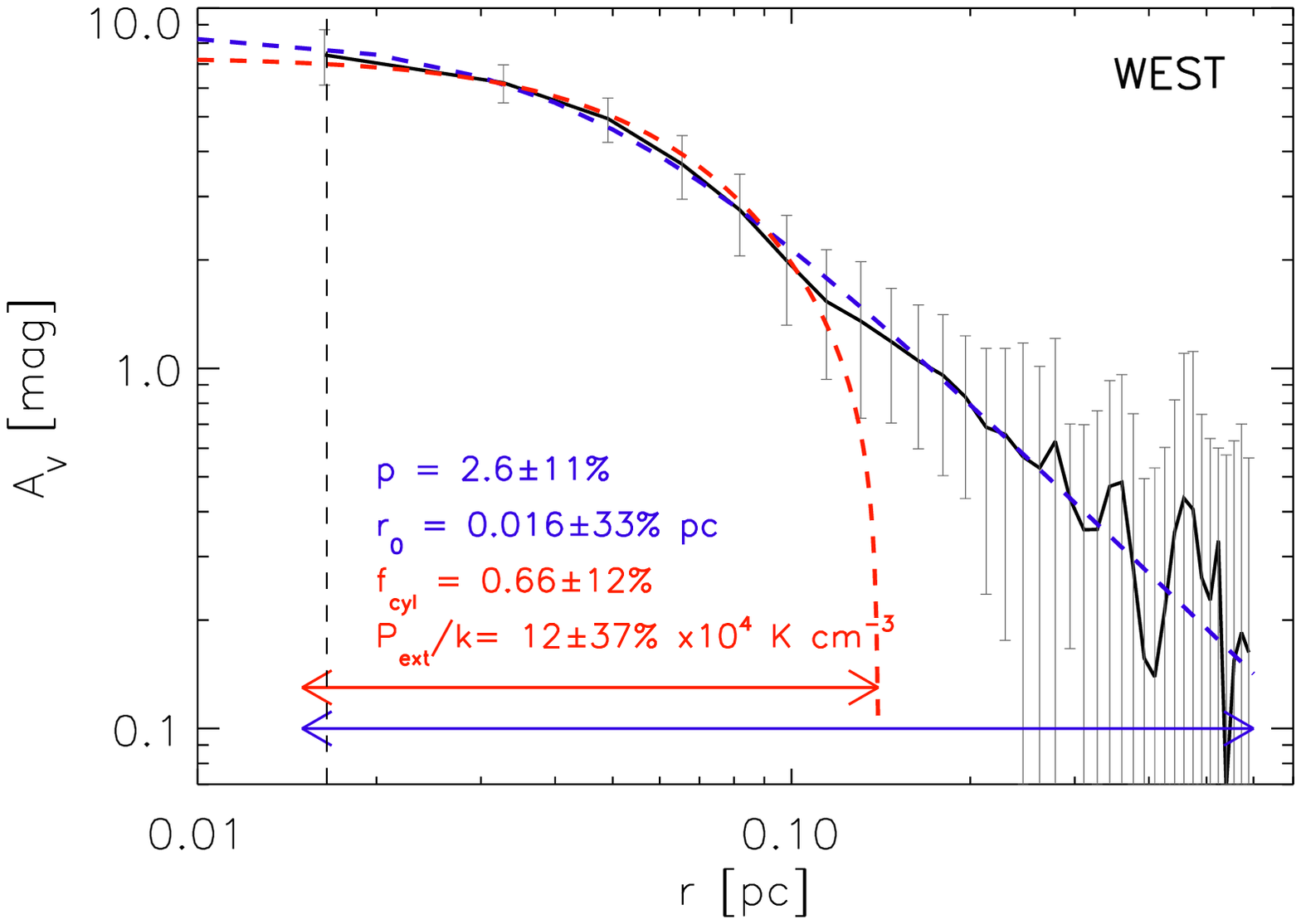}\includegraphics[width=\columnwidth]{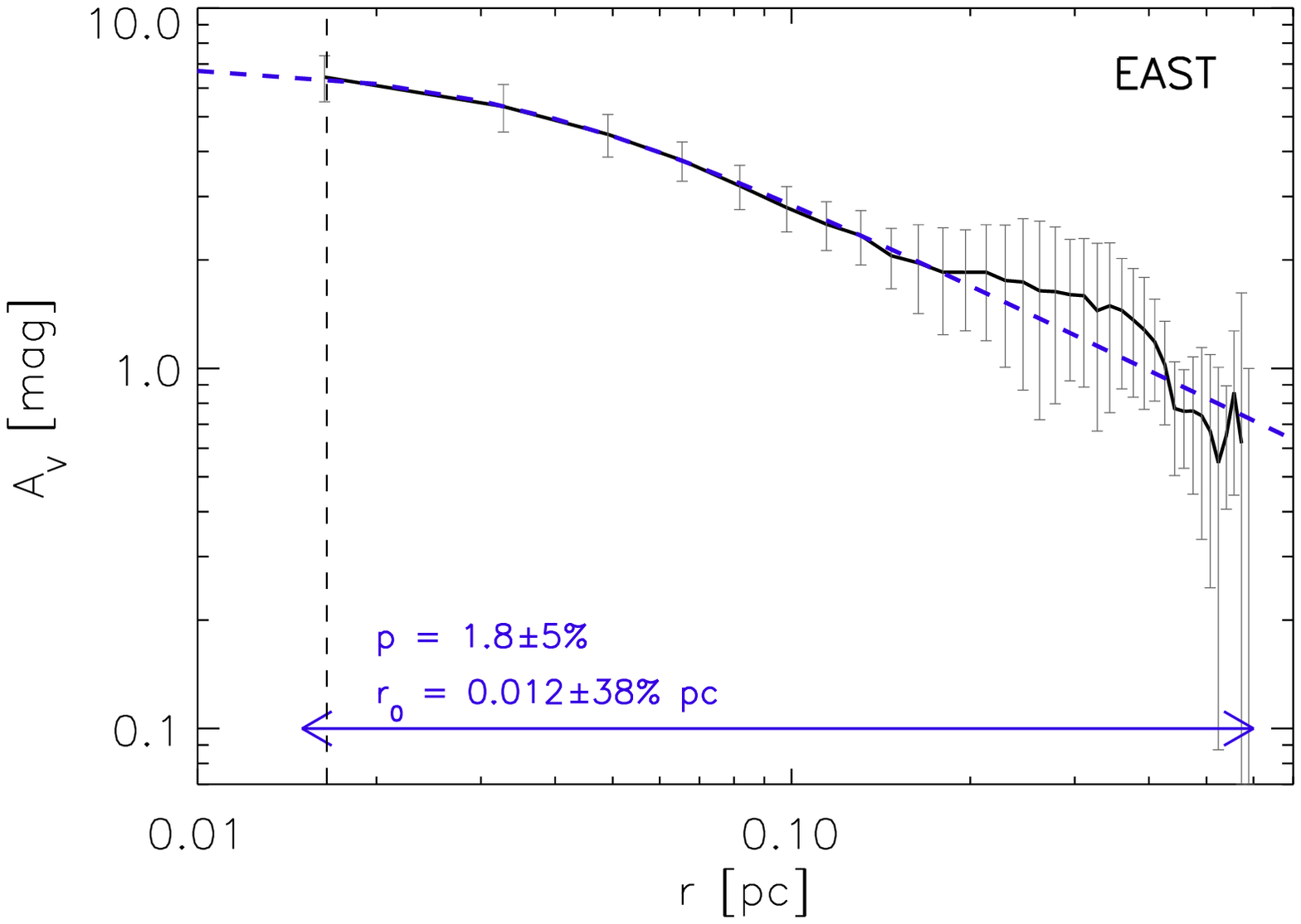}
      \caption{Mean radial column density profiles along the Center filament of the Musca cloud. The left and right panels show the mean profile on the west and east sides of the filament axis, respectively.
The blue line shows the fit of Plummer-like profile, and red line shows the the fit of a pressure-confined hydrostatic cylinder. The colored arrows indicate the radial ranges used in fitting the profiles. The error bars show the standard deviation of 20 individual profiles from which the mean profile is constructed. The vertical dashed line shows the $FWHM$ size of the extinction mapping beam. 
              }
         \label{fig:radialmodels}
   \end{figure*}

Figure \ref{fig:radialmodels} shows the best fit of Eq. \ref{eq:arz} to the observed mean profile, achieved by treating $\rho_\mathrm{c}$, $r_\mathrm{0}$, and $p$ as free parameters, fixing the fitting range to $r=[0, 0.6]$ pc, and then minimising the chi-square between the data points and the model. The inverse variances of the 20 individual profile cuts were used as the weights of the data points in the fit. The best-fit model for the west-side profile has the exponent $p = 2.6 \pm 11\%$, the scale radius $r_\mathrm{0} = 0.016 \pm 33\%$ pc, and the central density $n_\mathrm{c} = 54000 \pm 50\%$ cm$^{-3}$ (Table \ref{tab:fits}). The quoted errors are the one-sigma uncertainties of the fit. One should note that fitting Eq. \ref{eq:arz} in this manner suffers from degeneracy between $\rho_\mathrm{c}$, $r_\mathrm{0}$, and $p$ \citep{juv12a, smi14}. This degeneracy likely causes the large uncertainties in $r_\mathrm{0}$ and $\rho_\mathrm{c}$. The slope $p$ suffers less from the degeneracy and in general can be constrained with lower uncertainty \citep{smi14}.


\begin{table*}
\caption{Parameters of the best fit radial density distribution models.}   
\label{tab:fits}            
\centering                  
\begin{tabular}{l c c c c c c c l}
\hline\hline                
filament model 	& $\Delta r$\tablefootmark{a}		& $R_\mathrm{f}$	& $p$		& $r_\mathrm{0}$ 		& $n_\mathrm{c}$ 	& $f_\mathrm{cyl}$	& $P_\mathrm{ext}/k$	& other \\    
& 	[pc]			&  	[pc]		&		& [pc]		 		& [10$^4$ cm$^{-3}$]		& 				& [$10^4$ K cm$^{-3}$]& 	\\    
\hline                        
Plummer\tablefootmark{b}, west	&   $\leq$0.6		& -				& 2.6$\pm$11\%	&  0.016$\pm$33\%	& 5	   	& -				& - & - \\
Plummer\tablefootmark{b}, east	&   $\leq$0.6		& -				& 1.8$\pm$5\%	&  0.012$\pm$38\%	& 5	   	& -				& - & - \\
P-confined\tablefootmark{c}					&   $\leq$$R_\mathrm{f}$		& 0.12$\pm$7\%	& -			& -				& 	10			& 0.66$\pm$12\%	& 12$\pm$37\% & - 	\\
P-confined, with   		&   $\leq$0.6		& 0.10$\pm$12\%	& -			& -				&	10			& 0.64$\pm$18\% 	& 13$\pm$52\%  & $R_\mathrm{e}$=0.6 pc,  $N_\mathrm{0}$= \\     
\hspace{0.1 cm} envelope\tablefootmark{d}   		&  				& & 			& 				&				&  		&    &  2.4$\times$10$^{21}$ cm$^{-2}$	\\
 & & & & & & & & \hspace{0.2cm} \\     
\end{tabular}
\tablefoot{
\tablefoottext{a}{The fitting range.}
\tablefoottext{b}{Eq. \ref{eq:arz}.}
\tablefoottext{c}{Eq. \ref{eq:fischera}.}
\tablefoottext{d}{Eqs. \ref{eq:fischera} and \ref{eq:envelope2}.}}
\end{table*}

\subsubsection{Isothermal filament in equilibrium, confined by external pressure}
\label{subsec:radial_structure_embedded}


We also modelled the filament in the Center region as a pressure-confined (and truncated), isothermal filament in hydrostatic equilibrium \citep{fis12a}. The column density profile of this model is
\begin{eqnarray}
N (x) & = & \sqrt{\frac{p_\mathrm{ext}}{4 \pi G}}   \frac{\sqrt{8}}{1-f_\mathrm{cyl}} (\mu m_\mathrm{H})^{-1} \nonumber \\
			   &   &  \times \frac{1-f_\mathrm{cyl}}{1-f_\mathrm{cyl} + x^2f_\mathrm{cyl}} \bigg[  \sqrt{ f_\mathrm{cyl}(1-f_\mathrm{cyl})(1-x^2 )    }   \nonumber \\
			   &   & +  \sqrt{   \frac{1-f_\mathrm{cyl}}{1-f_\mathrm{cyl}(1-x^2)}      }      \nonumber \\
                               &     & \times   \arctan{   \sqrt{ \frac{f_\mathrm{cyl}(1-x^2)}{1-f_\mathrm{cyl}(1-x^2)}     }  }   \bigg].  
\label{eq:fischera}                               
 \end{eqnarray}
In the equation the dimensionless coordinate $x$ is the radius, $r$, normalized by the outer radius of the filament, $R_\mathrm{f}$, i.e., $x=r/R_\mathrm{f}$, and $f_\mathrm{cyl}=M_\mathrm{l} / M_\mathrm{l, crit}$ is the ratio of the filament's line mass to the critical line mass, (see Eq. \ref{eq:ostriker} for $M_\mathrm{l, crit}$). The relative shape of the column density profile depends only on the parameter $f_\mathrm{cyl}$ and on the normalisation of the radial profile, i.e., on $R_\mathrm{f}$. The temperature and line mass affect the profile shape implicitly through $f_\mathrm{cyl}$. In addition, the normalisation of the profile is affected by the confining external pressure $p_\mathrm{ext}$. 

The applicability of this model can be immediately examined through $f_\mathrm{cyl}$, which in the context of this model cannot be larger than unity. In Section \ref{subsec:cd_structure}, we estimated the line mass of the filament to be $M_\mathrm{l} = \{31, 26, 21\}$ M$_\odot$ when using the column density thresholds of $N_\mathrm{H} = \{1.5, 3, 4.5\} \times 10^{21}$ cm$^{-2}$ to compute the line mass. The critical line mass without the contribution from non-thermal motions (16 M$_\odot$ pc$^{-1}$) is slightly below this range, and the critical line mass with the non-thermal contribution (24 M$_\odot$ pc$^{-1}$) overlaps with it.  One should note that the observed line masses are upper limits because of the assumption of zero inclination. Distance uncertainty is also at least some 10\%. We also do not know exactly the temperature structure of the filament, which also affects $M_\mathrm{l, crit}$. Finally, in the context of this model, the filament is pressure-confined by the surrounding gas; this surrounding gas possibly contributes to the observed $M_\mathrm{l}$ value (we address this further below). We conclude that the observed line masses are such that considering the pressure-confined filament model is reasonable. 


We fitted Eq. \ref{eq:fischera} to the observed data points at radii below 0.6 pc by letting $f_\mathrm{cyl}$ and $p_\mathrm{ext}$ change as free parameters, again using the inverse variances of the 20 profiles to weight the data points. This did not yield an acceptable fit. However, allowing also the fitting radius, i.e., the radius of the filament, to change as a free parameter and ignoring the profile outside it resulted in a reasonable fit (see Fig. \ref{fig:radialmodels}). The best fit has the parameters $f_\mathrm{cyl} = 0.66 \pm 12\%$, and $p_\mathrm{ext} = 12 \pm 37\% \times 10^4$ K cm$^{-3}$, and $R_\mathrm{f} = 0.12 \pm 7\%$ pc where the uncertainties result from the chi-square fit (Table \ref{tab:fits}). The external pressure is consistent with pressures needed to confine isolated globules modelled as Bonnor-Ebert spheres \citep[$\sim$2-20$\times 10^4$ K cm$^{-3}$,][]{kan05}, and a factor of a couple higher than inferred for parsec-scale filaments \citep[$\sim$1-5$\times 10^4$ K cm$^{-3}$,][]{fis12a, fis12b}.

One has to recognise that the above model does not include any contribution to the observed column densities from the surrounding gas that is required to provide the confining pressure \citep[see the discussion in][]{rec14}. Thus, the use of Eq. \ref{eq:fischera} by itself may not be physically entirely meaningful.
%
%
In practice, the confining pressure could arise from the thermal pressure in a two-phase medium, in which the cooler, denser component is surrounded (and confined) by the less dense, warm component. Accretion of gas into the filament could also possibly provide confining pressure \citep{hei13a, hei13b}. Turbulent ram pressure could also provide confining pressure, although to provide confinement, it should be relatively isotropic, which may be unrealistic. Regardless of the nature of the pressure, a surrounding gas component is needed, and it can affect the observed column density structure. 


   \begin{figure}
   \centering
\includegraphics[width=\columnwidth]{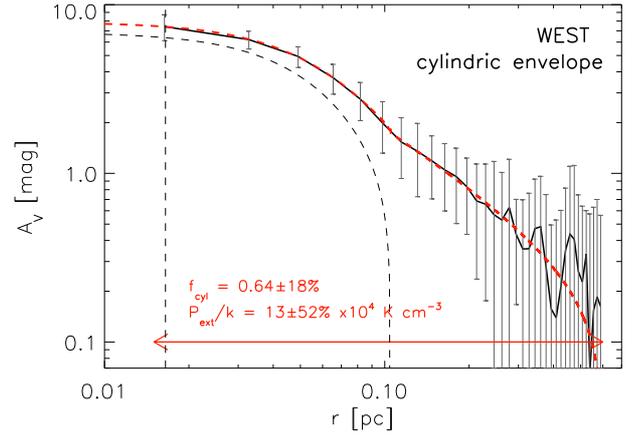}
      \caption{Mean radial column density profiles along the Center filament. The red line shows the fit of a pressure-confined equilibrium model \citep{fis12a}, embedded inside a cylindric envelope. The black dashed line shows the column density profile of the filament without the surrounding envelope. The dashed vertical line shows the $FWHM$ size of extinction mapping beam.
              }
         \label{fig:cartoon}
   \end{figure}


To explore the effect of the surrounding gas to the parameters of this model, we made an experiment in which the pressure-confined filament is embedded inside a cylindric envelope \citep[e.g.,][]{her12, rec14}. This cylindric envelope has a power-law radial density distribution, $\rho_\mathrm{e} (r) = C r^{-2}$, where $C$ is a proportionality constant. The column density profile of the envelope results then from integrating $\rho_\mathrm{e}(r)$ along the line of sight \citep[see, e.g., ][]{rec14}, yielding
\begin{equation}
N_\mathrm{e}(r) = \frac{C}{r} 
\begin{cases}
\arctan{\sqrt{ (R_\mathrm{e}/r)^2 - 1}}    & \\
\hspace{24pt} - \arctan{\sqrt{(R_\mathrm{f}/r)^2 - 1}}, & r < R_\mathrm{f} \\
\arctan{\sqrt{ (R_\mathrm{e}/r)^2 - 1}},     & R_\mathrm{f} < r < R_\mathrm{e},
\end{cases}
\label{eq:envelope2}
\end{equation} 	
where $R_\mathrm{e}$ and $R_\mathrm{f}$ are the radius of the envelope and filament, respectively. At the limit $r \rightarrow 0$ the equation has the value $C(-R_\mathrm{e}^{-1} + R_\mathrm{f}^{-1})$, which gives the column density of the envelope at $r=0$, i.e., $N_\mathrm{0}$. The total column density of the model is the sum of the envelope and the embedded filament (Eq. \ref{eq:fischera}). This model was fitted to the observed profile with $f_\mathrm{cyl}$, $p_\mathrm{ext}$, $R_\mathrm{f}$, and $C$ as free parameters, assuming that the envelope goes to zero at 0.6 pc, i.e., $R_\mathrm{e} = 0.6$ pc (Fig. \ref{fig:cartoon}). We stress here that our goal is to probe how the filament parameters are affected by an envelope, not to find the best possible envelope model, nor even a physically meaningful one. It is evident that the envelope model we use is rather arbitrary, and other envelopes may well produce equally good fits.


We find that the observed profile is best fit when a filament is surrounded by an envelope that has the central column density $N_\mathrm{0} = 2.4 \times 10^{21}$ cm$^{-2}$. The parameters of the embedded filament, i.e., $f_\mathrm{cyl}$ and $p_\mathrm{ext}$ (see Table \ref{tab:fits}) are not significantly different from those of the filament without the envelope. In conclusion, a pressure-confined hydrostatic filament, embedded in a low-density envelope, seems to provide a reasonable explanation for the entire observed column density profile. 

\subsection{Relationship between the column density and velocity structure in Musca}
\label{subsec:velocity_structure}

%
%
We briefly compare in this section the column density structure of the Musca cloud to the large-scale kinematic data from \citet[][]{hac15}. While Hacar et al. discusses the properties of the global velocity field in Musca, we put here those data in the context of the column density structure and fragments of the cloud. 

\begin{figure*}
   \centering
       \includegraphics[bb = 10 120 555 680, angle=90, clip=true, width=\textwidth]{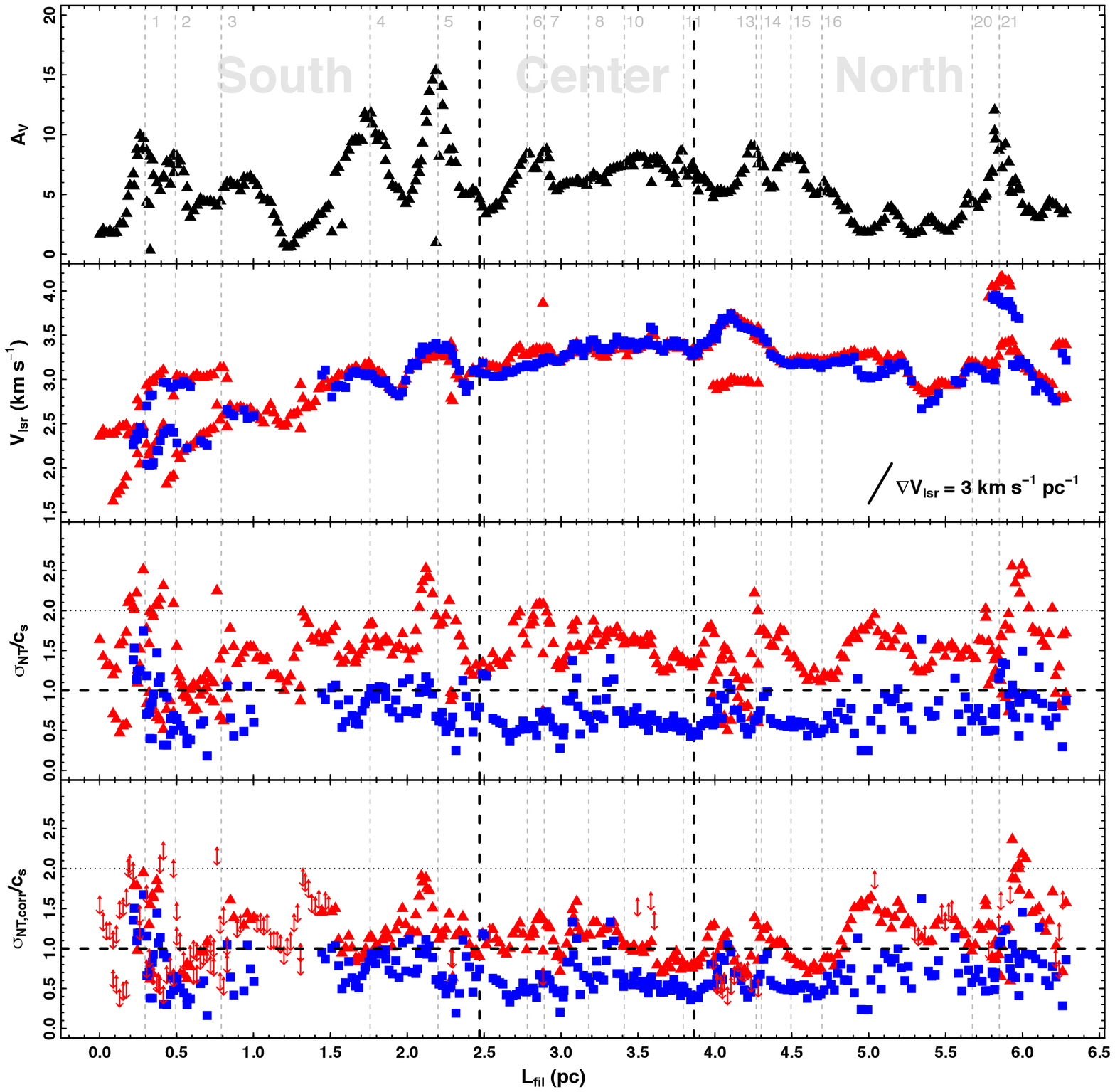}   
 \caption{Column density and velocity structure of Musca, measured approximately along the crest of the cloud (see Fig. \ref{fig:musca_large}). The top panel shows the dust extinction from NIR data. The second panel shows the central velocities, $V_\mathrm{lsr}$. The third and bottom panels show the non-thermal velocity dispersions, $\sigma_\mathrm{NT}$ (non-corrected) and $\sigma_\mathrm{NT, corr}$ (opacity-corrected), normalised by the sound speed $c_\mathrm{s}$. The arrows in the bottom panel indicate upper limits. The red and blue points show the data for $^{13}$CO (2-1) and C$^{18}$O (2-1), respectively. The locations of the fragments that are closer than 100 arcsec of the cloud crest are indicated with thin dashed lines. The thick dashed vertical lines separate from each others the South, Center, and North regions. The thick dashed horizontal line indicates the sonic velocity dispersion and the dotted horizontal line the two times the sonic velocity dispersion. The lower right corner of the middle panel indicates with a line a velocity gradient of 3 km s$^{-1}$. The figure is a reproduction from \citet{hac15}. 
              }
         \label{fig:velocities}
   \end{figure*}

Figure \ref{fig:velocities} shows the NIR-derived column densities, $^{13}$CO and C$^{18}$O central velocities ($V_\mathrm{lsr}$), non-thermal velocity dispersions ($\sigma_\mathrm{NT}$), and non-thermal opacity-corrected velocity dispersions ($\sigma_\mathrm{NT, corr}$) for the $\sim$300 line-of-sights observed along the Musca cloud (see Fig. \ref{fig:musca_large}). Note there are some double-peaked spectra at the edges of the cloud; Fig. \ref{fig:velocities} shows the $V_\mathrm{lsr}$ and $\sigma_\mathrm{NT}$ values for both these components, obtained via fits of separate Gaussian components \citep[see][for details]{hac15}. The locations of the fragments that are closer than 100 arcsec of the cloud crest are also shown in the figure. As found out by \citet{hac15}, Musca is velocity-coherent and subsonic at densities probed by the C$^{18}$O data (sub-/transsonic at densities probed by the $^{13}$CO data) practically throughout its length. The velocity gradient along the South region is about $\mathrm{d}V_\mathrm{lsr} / \mathrm{d}r \approx 0.5$ km s$^{-1}$ pc$^{-1}$. The South region also shows local velocity gradients, up to $\sim$3 km s$^{-1}$ \citep[cf.,][]{hac15}, that clearly coincide with the column density fragments. The North and Center regions show a velocity gradient of $\mathrm{d}V_\mathrm{lsr} / \mathrm{d}r \approx 0.15$ km s$^{-1}$ pc$^{-1}$, with the only local gradients coinciding with the densest fragment in the North region (\#21) and the complex clump of multiple maxima (close to fragment \#13). In particular, the two most pronounced fragments in the Center region (\#6 and 7 in Table \ref{tab:cores}) do not show gradients in C$^{18}$O central velocities. The scatter of the line velocities in The North and South is clearly larger than in the Center (dominantly due to the large-scale gradient): $\sigma(V_\mathrm{LSR}, \mathrm{C}^{18}\mathrm{O})_\mathrm{south} = 2.03$ km s$^{-1}$, $\sigma(V_\mathrm{LSR}, \mathrm{C}^{18}\mathrm{O})_\mathrm{centre} = 0.67$ km s$^{-1}$, $\sigma(V_\mathrm{LSR}, \mathrm{C}^{18}\mathrm{O})_\mathrm{north} = 1.57$ km s$^{-1}$. The North and South  also show slightly larger scatter of non-thermal velocity dispersions than the Center: $\sigma(\sigma_\mathrm{NT}, \mathrm{C}^{18}\mathrm{O})_\mathrm{south} = 0.3$ km s$^{-1}$, $\sigma(\sigma_\mathrm{NT}, \mathrm{C}^{18}\mathrm{O})_\mathrm{centre} = 0.2$ km s$^{-1}$, $\sigma(\sigma_\mathrm{NT}, \mathrm{C}^{18}\mathrm{O})_\mathrm{north} = 0.24$ km s$^{-1}$. Finally, some  oscillatory motions are seen at the locations of the two most prominent fragments in the South region (\#4 and 5). Their presence suggests that fragmenting modes indeed are present in the cloud \citep[cf.,][]{hac11}. 

Summarising, the above properties demonstrate the fact that even if Musca as a whole is subsonic and velocity-coherent, the opposite ends of it show slightly more active internal dynamics than the extremely quiescent Center filament. This activity is likely related to the fragments present in the ends of the cloud.


\section{Discussion}  
\label{sec:discussion}

\subsection{Musca: caught in the middle of gravitational fragmentation}

What do our observations tell about the nature of Musca and its fragmentation? We have demonstrated that the physical structure of the extremely quiescent Center filament of Musca is in a good agreement with the predictions for the line mass, radial density distribution, and velocity structure of the pressure-confined hydrostatic equilibrium model. The cloud contains 21 fragments (according to our definition of fragments), with the higher-contrast ones located in the South and North regions. Specifically, the North and South have a factor of three higher column density variance compared to the Center region. Some of the fragments contain several solar masses, and thus have the potential to be pre-stellar cores. Importantly, the equivalent line masses of the North and South are equal with the line mass of the Center filament, giving rise to a hypothesis that \emph{they may have fragmented from an initial condition that resembles the relatively non-structured Center filament}. 

To further examine this hypothesis, we consider the prediction of the gravitational fragmentation model \citep[the pressure-confined filament,][]{fis12a} for the fragmentation length and time-scale of a cylinder that has the properties of the Center filament. In this model, filaments with line masses less than the critical line mass can fragment when perturbed. Filaments with line masses larger than critical (that are not equilibrium solutions to begin with) should collapse into a spindle rapidly. The wavelength of the fastest growing perturbations is between $\lambda_\mathrm{m} = 5-6.3 \times FWHM_\mathrm{cyl}$ \citep{fis12a}. The lower limit corresponds to the case $f_\mathrm{cyl} \rightarrow 1$ and the upper limit to $f_\mathrm{cyl} \rightarrow 0$. We measured the $FWHM$ of the Center filament to be 0.07 pc, which results in the fragmentation scale of $\lambda_\mathrm{m} \approx 0.35-0.44$ pc. The observed mean peak separation of the fragments in all regions is $0.41 \pm 0.37$ pc and the median is 0.34 pc (mean is $0.32 \pm 0.18$ pc and median 0.33 pc without the long separation between fragments \#3 and 4), overlapping with the predicted interval. The results are practically the same regardless of how we compute the separations in the North region (see Section \ref{subsec:cd_structure}). We also note that fragment \#9 is not included in this analysis because of its location off the filament axis. In conclusion, it appears that also from the fragmentation scale point-of-view, the North and South regions may have originated from an initial condition that resembles the Center filament. 

The fragmentation time-scale of the infinite, pressure-confined cylinder depends on $f_\mathrm{cyl}$, $T$, and $p_\mathrm{ext}$ \citep{fis12a}, and is between 
\begin{equation}
\tau_\mathrm{m} = 
\begin{cases}
2.95 (1-f_\mathrm{cyl})\tau_\mathrm{Kp}    	& ,  f_\mathrm{cyl} \rightarrow 1 \\
4.08 \tau_\mathrm{Kp}    					& , f_\mathrm{cyl} \rightarrow 0, \\
\end{cases}
\end{equation} 
where $\tau_\mathrm{Kp} = \sqrt{kT/ (4\pi \mu m_\mathrm{H}G p_\mathrm{ext})} \approx 0.2$ Myr. This gives the range $\tau_\mathrm{m}\approx 0.2-0.8$ Myr when using the parameters of the best-fitting pressure-confined model (Table \ref{tab:fits}). We note the caveat that the above predictions for the fragmentation scale and time-scale are for infinite cylinders; it may be that the predictions differ in finite geometry.

Why has the fragmentation proceeded strongly at the ends of the Musca cloud, but not at its center? One possible mechanism in the context of gravitational fragmentation is provided by finite-size effects in cylindric geometry \citep[as opposed to the infinite models of \citealt{ost64} and][]{fis12a}. In finite filaments, the acceleration of gas depends on the relative position along the filament, being largest at the ends of it \citep[][see also \citealt{bur04} for a study in a sheet geometry]{bas83, pon11, pon12, cla15}. These higher accelerations may translate into shorter local collapse time-scales, and therefore, into formation of fragments at the ends of the filament prior to its centre \citep{pon11, pon12}. For example, \citet[][see their Figs. 5 and 6]{pon11} suggests that the local collapse might occur a factor of two-to-three faster at the ends of the filament than at its center. 
%
%
Similarly, \citet[][]{cla15} found in their numerical simulations that the first fragments form at the ends of the filament, regardless of its aspect ratio. While there are some quantitative differences between the predictions of \citet{cla15} and \citet{pon11, pon12}, both models predict that long (but finite-sized) filaments fragment starting from their ends. Consequently, they lay out the basis for explaining the fragmentation at the ends of Musca as a result of the end-dominated collapse caused by higher accelerations, and therefore, shorter local collapse time-scales.  

In the models of \citet{pon11, pon12} and \citet{cla15}, high-aspect ratio filaments collapse globally and the fragments formed at the ends of the cloud fall into its centre; this can be used to gain further insight into the evolution of Musca under this framework. The time-scales for the global collapse (time-scale for the end fragments to reach the cloud centre) are \citep[following the notation of][]{cla15}
\begin{equation}
t_\mathrm{cla} = (0.49 + 0.26 \ A) \ (G\rho_\mathrm{0})^{-1/2}  , 
\end{equation}
and
\begin{equation}
t_\mathrm{pon} = 
\begin{cases}
0.44 \ A \ (G\rho_\mathrm{0})^{-1/2}           \quad\quad\quad\quad\quad & A \lesssim 5 \\
0.98 \ A^\mathrm{1/2} \ (G\rho_\mathrm{0})^{-1/2}        & A \gtrsim 5,
\end{cases}
\label{eq:timescales}
\end{equation} 
where G is the gravitational constant and $\rho_\mathrm{0}$ the central density. The aspect ratio of Musca depends on how we define its width. If we adopt the $\sim$0.6 pc extent of the Plummer profile (see Fig. \ref{fig:radialmodels}), the aspect ratio 6.5 pc / 0.6 pc $\approx$ 5 follows; if we adopt the 0.1 pc radius of the pressure-confined filament, the aspect ratio is 6.5 pc / 0.2 pc $\approx$ 32. This interval of $A=5-33$ is very conservative, as it represents the extreme choices for the filament width. For the central density, we use $10^5$ cm$^{-3}$. The global collapse time-scale for the chosen range of aspect ratios is then between $\sim$0.5-3 Myr, depending on the model. We showed earlier that the fragmentation time-scale of an infinite pressure-confined filament is between 0.2-0.8 Myr. Even if these estimates are highly uncertain, they suggest that the fragmentation time-scale may be significantly shorter than the global collapse time-scale. This is in agreement with the fact that we do not observe Musca to be globally collapsing, but rather forming multiple fragments along it. This, in turn, suggests that the age of Musca is rather closer to its fragmentation time-scale than its global collapse time-scale, i.e., only some tenths of a Myr. Alternatively, additional processes must be supporting Musca against global collapse.


Instead of the gravitational focusing, another possibility to explain the differences in fragmentation along Musca may be differences in the initial physical properties along it. In principle, the fragmentation time-scale anti-correlates with the line mass \citep[e.g.,][]{fis12a}; however as shown earlier, the equivalent line masses of the North and South are the same as the line mass of the Center, stating that they should collapse in the same time-scale. It could also be hypothesised that difference in the dynamics of the regions plays a role: the North and South regions show significantly higher global velocity gradients than the Center region, which might affect fragmentation in them. Whether such an initial velocity gradient could promote fragmentation remains to be explored via theoretical works. Finally, the time-scale also correlates with the size of the initial density perturbations \citep[][]{pon11}. It cannot be ruled out that the North and South regions would have had higher density variance along them compared to the Center, inducing faster fragmentation. 

Ultimately, one would like to compare the time-scales derived above with other age estimators in the cloud. However, the only direct sign-post of age in Musca is the single T Tauri star candidate in the North region (fragment \#21). Its presence indicates that fragmentation in the North region (and supposedly in South) must have taken place on the order of $\sim$0.5 Myr ago \citep[Class 0+1 protostellar lifetime,][]{dun14}, with an admittedly large uncertainty. We showed earlier that the time-scale of gravitational fragmentation in Musca is 0.2-0.8 Myr, coinciding with this estimate. Further studies targeting, e.g., the chemical ages at the different locations along the filament would provide additional, independent measures of age.

Altogether, the morphology of Musca, its fragmentation, radial structure, and velocity characteristics suggest that gravitational fragmentation is taking place \emph{in situ} in Musca, moulding it towards star formation. This is in a qualitative agreement with our earlier results regarding the cloud's column density structure: in \citet{kai09} we analysed the probability density function of column densities ($N$-PDF) in Musca, showing that it exhibits (at least a start of) a power-law-like tail at high-column densities, similarly with actively star-forming clouds. Further, in \citet{kai14} we showed with the help of volume density modelling that Musca currently contains only a very small amount of gas ($\sim$6 M$_\odot$) at densities high-enough for star formation. From the point-of-view of its column and volume density structure, the cloud must become more condensed locally to form stars; this could happen via the on-going fragmentation and gravitational collapse. Musca represents a perfect target for studying this: the simplicity of the cloud is ideal for isolating the different processes driving fragmentation. 

\section{Conclusions} 
\label{sec:conclusions}

We analyzed in this work the column density structure and fragmentation of the 6.5 pc long, rectilinear Musca molecular cloud that shows very little signs of on-going star formation. We examined the structure of the cloud with the help of near-infrared dust extinction data and 870 $\mu$m dust emission data. We also examined the large-scale kinematic structure of Musca using $^{13}$CO and C$^{18}$O line emission data \citep[from][]{hac15}. Our conclusions are as follows.
	
\begin{enumerate}

   \item The column density data show that the Musca cloud is fragmented into 21 fragments that have masses between $\sim$0.3-7 M$_\odot$. The Center region of Musca harbours a remarkably well-defined, 1.6 pc long filament that contains only low-contrast fragments. The filament shows only 15\% relative column density variations along its crest, in stark contrast to $\sim$50\% variations in the more pronouncedly fragmented ends of the cloud. We emphasize that the relative column density variations provide a physically motivated, simple measure of how fragmented a filament is; a low degree of fragmentation is necessary for characterising the initial condition of fragmenting filaments. 
   
   \item The line mass of the Center filament is 21-31 M$_\odot$ pc$^{-1}$, i.e., close to the critical value of a cylinder in hydrostatic equilibrium. The equivalent line masses (total mass divided by length) of the North and South regions are practically the same. 
   
    \item The radial density structure of the Center filament is well described with either \emph{i)} a Plummer profile that has a power-law exponent of 2.6, scale radius of 0.016 pc ($FWHM$ of 0.07 pc), and central density of 5 $\times 10^4$ cm$^{-3}$, or \emph{ii)} a hydrostatic filament that is confined by external pressure imposed by a surrounding envelope. In this case the ratio of the line mass to the critical line mass is 0.66, central density is $10^{5}$ cm$^{-3}$, and external pressure is $\sim$10$^{5}$ K cm$^{-3}$. We note that even this simple and well-resolved filament shows an asymmetric column density profile with respect to the main filament axis. This advices caution when analysing profiles of filaments that are more complicated than Musca.
   
   \item The fragmentation length of the cloud, $\sim$0.4 pc, agrees with the prediction of the gravitational fragmentation of an infinite, pressure-confined cylinder. The fragmentation time-scale predicted by the model is 0.2-0.8 Myr and the global collapse time of the cloud 0.5-3 Myr.  

    \item The ends of Musca show somewhat more variation in their C$^{18}$O velocity structure than the extremely quiescent Center region, both in the central line velocities and the scatter of non-thermal line widths. The central line velocities peak at the locations of the fragments, suggesting motions associated with the fragments.
            
\item The above characteristics give rise to a hypothesis for the evolutionary scenario of Musca: an initially 6.5 pc long, close-to hydrostatic filament is in the middle of gravitational fragmentation. Fragmentation started a few tenths of a Myr ago, and it has so far taken place only at the ends of the cloud, possibly because of the shorter local collapse time-scales compared to the cloud centre (``gravitational focusing"). The relatively non-fragmented Center region may still represent the close-to hydrostatic initial condition of the filament, regardless of whether the fragmentation at the ends of the cloud is driven by gravitational focusing or some other process. 
                        
\end{enumerate}


\begin{acknowledgements}
The authors thank S. Recchi and F. Heitsch for valuable discussions, and the anonymous referee for constructive comments that improved the paper. 
The work of J.K. was supported by the Deutsche Forschungsgemeinschaft priority program 1573 ("Physics of the Interstellar Medium"). J.K. gratefully acknowledges support from the Finnish Academy of Science and Letters/V\"ais\"al\"a Foundation. H. Bouy is funded by the the Ram\'on y Cajal fellowship program number RYC-2009-04497. This research has been funded by Spanish grants AYA2012-38897-C02-01.
This publication makes use of data products from the Two Micron All Sky Survey, which is a joint project of the University of Massachusetts and the Infrared Processing and Analysis Center/California Institute of Technology, funded by the National Aeronautics and Space Administration and the National Science Foundation.
\end{acknowledgements}



\appendix

\section{Relationship between extinction-derived and emission-derived column densities}
\label{app:laboca-vs-nir}

We measured the column densities through Musca using two techniques: NIR dust extinction mapping and 870 $\mu$m continuum emission. We briefly compare here the column densities resulting from them. The extinction was used to estimate the column densities as explained in Section \ref{subsec:nirdata}. To estimate the column densities from emission, we used the formula:
\begin{equation}
N_\mathrm{LABOCA} = \frac{RF_\mathrm{\nu}}{ B_\mathrm{\nu}(T_\mathrm{dust})\mu m_\mathrm{H}\kappa \Omega},
\label{eq:laboca_N}
\end{equation}
where $F_\mathrm{\nu}$ is the flux density, $R=100$ is the gas-to-dust ratio, $B_\mathrm{\nu}$ is the Planck function in temperature $T_\mathrm{dust}=10$ K, $\mu=2.8$ is the mean molecular weight, $\kappa = 1.85$ cm$^{2}$ g$^{-1}$ the dust opacity \citep{oss94}, and $\Omega$ the beam solid angle. 

\begin{figure}
   \centering
\includegraphics[width=0.5\textwidth]{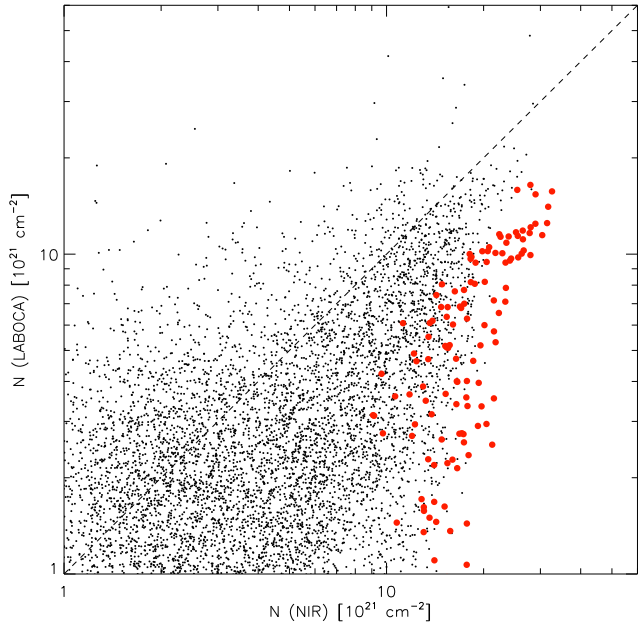}
      \caption{Relationship between emission-derived ($N$(LABOCA)) and extinction-derived ($N$(NIR)) column densities. The red symbols show the data in the area of the most massive fragment in the cloud (\#5). The dashed line shows the one-to-one relationship.}
         \label{fig_laboca-vs-nir}
   \end{figure}

Figure \ref{fig_laboca-vs-nir} show the relationship between the column densities from the two techniques. The relationship has a large scatter, likely resulting from the spatial filtering of the LABOCA data. The relationship indicates that no significant column density peaks remain undetected by the NIR data; there is no significant saturation of the NIR-derived column densities compared to LABOCA-derived ones. We illustrate this further by highlighting in Fig. \ref{fig_laboca-vs-nir} the relationship in the area of the most massive fragment (\#5) of the cloud. For this particular fragment, the NIR-derived column densities are systematically higher than the LABOCA-derived ones. This could be, e.g., due to misestimation of the temperature used in Eq. \ref{eq:laboca_N}.

\section{Definition of the Center filament crest}
\label{app:crest}

To measure the radial profile of the Center filament, we defined its approximate crest with lines connecting nine points. The coordinates of these points are given in Table \ref{tab:crest}. This crest was sampled with 20 equally-spaced cuts perpendicular to it. 


\begin{table}
\caption{The points defining the crest of the Center filament}             
\label{tab:crest}     
\centering                    
\begin{tabular}{c c}     
\hline\hline            
R.A. 	& Dec.   \\   
\hline            
   12:29:52.5 	&   -71:07:19 \\
   12:29:37.9 	&   -71:09:28 \\
   12:29:22.9	&   -71:10:49 \\
   12:29:00.5	&   -71:14:06 \\
   12:28:42.3	&   -71:17:17 \\
   12:27:47.5	&   -71:21:08 \\
   12:27:02.3	&   -71:27:26 \\
   12:26:46.2	&   -71:29:48 \\
   12:26:20.7	&   -71:33:20 \\
\hline               
\end{tabular}
\end{table}


\begin{thebibliography}{}



\bibitem[Andr{\'e} et al.(2014)]{and14} Andr{\'e}, P., Di Francesco, J., Ward-Thompson, D., et al.\ 2014, Protostars and Planets VI, 27 

\bibitem[Andr{\'e} et al.(2010)]{and10} Andr{\'e}, P., Men'shchikov, A., Bontemps, S., et al.\ 2010, \aap, 518, L102 

\bibitem[Arzoumanian et al.(2011)]{arz11} Arzoumanian, D., Andr{\'e}, P., Didelon, P., et al.\ 2011, \aap, 529, L6 

\bibitem[Autry et al.(2003)]{2003SPIE.4841..525A} Autry, R.~G., Probst, 
R.~G., Starr, B.~M., et al.\ 2003, \procspie, 4841, 525 

\bibitem[Bally et al.(1987)]{bal87} Bally, J., Lanber, W.~D., Stark, A.~A., \& Wilson, R.~W.\ 1987, \apjl, 312, L45 

\bibitem[Bastien(1983)]{bas83} Bastien, P.\ 1983, \aap, 119, 109 

\bibitem[Barnard(1927)]{bar27} Barnard, E.~E. A Photographic Atlas of Selected Regions of the Milky Way. Ed. Edwin B. Frost and Mary R. Calvert. Washington: Carnegie Institution of Washington, 1927

\bibitem[Belloche et al.(2011)]{BEL11} Belloche, A., Schuller, F., Parise, B., et al.\ 2011, \aap, 527, A145 

\bibitem[Bergin \& Tafalla(2007)]{ber07} Bergin, E.~A., \& Tafalla, M.\ 2007, \araa, 45, 339 

\bibitem[Bertin(2011)]{PSFEx} Bertin, E.\ 2011, Astronomical Data Analysis Software and Systems XX, 442, 435 

\bibitem[Bertin(2006)]{2006ASPC..351..112B} Bertin, E.\ 2006, Astronomical Data Analysis Software and Systems XV, 351, 112 

\bibitem[Bertin \& Arnouts(1996)]{1996A&AS..117..393B} Bertin, E., \& Arnouts, S.\ 1996, \aaps, 117, 393 

\bibitem[Bertin et al.(2002)]{2002ASPC..281..228B} Bertin, E., Mellier, Y., Radovich, M., et al.\ 2002, Astronomical Data Analysis Software and Systems XI, 281, 228 

\bibitem[Bohlin et al.(1978)]{boh78} Bohlin, R.~C., Savage, B.~D., \& Drake, J.~F.\ 1978, \apj, 224, 132 

\bibitem[Burkert \& Hartmann(2004)]{bur04} Burkert, A., \& Hartmann, L.\ 2004, \apj, 616, 288 

\bibitem[Busquet et al.(2013)]{bus13} Busquet, G., Zhang, Q., Palau, A., et al.\ 2013, \apjl, 764, L26 

\bibitem[Cardelli et al.(1989)]{car89} Cardelli, J.~A., Clayton, G.~C., \& Mathis, J.~S.\ 1989, \apj, 345, 245 

\bibitem[Clarke \& Whitworth(2015)]{cla15} Clarke, S.~D., \& Whitworth, A.~P.\ 2015, arXiv:1502.07552 

\bibitem[Dobashi et al.(2005)]{dob05} Dobashi, K., Uehara, H., Kandori, R., et al.\ 2005, \pasj, 57, 1 

\bibitem[Dunham et al.(2014)]{dun14} Dunham, M.~M., Stutz, A.~M., Allen, L.~E., et al.\ 2014, Protostars and Planets VI, 195 

\bibitem[Fiege \& Pudritz(2000a)]{fie00a} Fiege, J.~D., \& Pudritz, R.~E.\ 2000a, \mnras, 311, 85 

\bibitem[Fiege \& Pudritz(2000b)]{fie00b} Fiege, J.~D., \& Pudritz, R.~E.\ 2000b, \mnras, 311, 105 

\bibitem[Fischera \& Martin(2012a)]{fis12a} Fischera, J., \& Martin, P.~G.\ 2012a, \aap, 542, A77  

\bibitem[Fischera \& Martin(2012b)]{fis12b} Fischera, J., \& Martin, P.~G.\ 2012b, \aap, 547, A86 



\bibitem[Goodman et al.(2014)]{goo14} Goodman, A.~A., Alves, J., Beaumont, C.~N., et al.\ 2014, \apj, 797, 53 

\bibitem[G{\"u}ver \& \"Ozel(2009)]{gue09} G{\"u}ver, T., \& \"Ozel, F.\ 2009, \mnras, 400, 2050 

\bibitem[Hacar \& Tafalla(2011)]{hac11} Hacar, A., \& Tafalla, M.\ 2011, \aap, 533, A34 

\bibitem[Hacar et al.(submitted)]{hac15} Hacar, A., Kainulainen, J., Beuther, H., Tafalla, M., Alves, J.\ submitted to A\&A. 

\bibitem[Hacar et al.(2013)]{hac13} Hacar, A., Tafalla, M., Kauffmann, J., \& Kov{\'a}cs, A.\ 2013, \aap, 554, A55 


\bibitem[Hernandez et al.(2012)]{her12} Hernandez, A.~K., Tan, J.~C., Kainulainen, J., et al.\ 2012, \apjl, 756, L13 

\bibitem[Heitsch(2013a)]{hei13a} Heitsch, F.\ 2013a, \apj, 769, 115 

\bibitem[Heitsch(2013b)]{hei13b} Heitsch, F.\ 2013b, \apj, 776, 62 

\bibitem[Hill et al.(2011)]{hil11} Hill, T., Motte, F., Didelon, P., et al.\ 2011, \aap, 533, A94 



\bibitem[Inutsuka \& Miyama(1992)]{inu92} Inutsuka, S.-I., \& Miyama, S.~M.\ 1992, \apj, 388, 392 

\bibitem[Inutsuka \& Miyama(1997)]{inu97} Inutsuka, S.-i., \& Miyama, S.~M.\ 1997, \apj, 480, 681 

\bibitem[Jackson et al.(2010)]{jac10} Jackson, J.~M., Finn, S.~C., Chambers, E.~T., Rathborne, J.~M., \& Simon, R.\ 2010, \apjl, 719, L185 

\bibitem[Juvela et al.(2012a)]{juv12a} Juvela, M., Malinen, J., \& Lunttila, T.\ 2012a, \aap, 544, A141 
\bibitem[Juvela et al.(2012b)]{juv12b} Juvela, M., Ristorcelli, I., Pagani, L., et al.\ 2012b, \aap, 541, A12    

\bibitem[Juvela et al.(2011)]{juv11} Juvela, M., Ristorcelli, I., Pelkonen, V.-M., et al.\ 2011, \aap, 527, A111 

\bibitem[Juvela et al.(2010)]{juv10} Juvela, M., Ristorcelli, I., Montier, L.~A., et al.\ 2010, \aap, 518, L93 

\bibitem[Kainulainen et al.(2014)]{kai14} Kainulainen, J., Federrath, C., \& Henning, T.\ 2014, Science, 344, 183 

\bibitem[Kainulainen et al.(2013)]{kai13} Kainulainen, J., Ragan, S.~E., Henning, T., \& Stutz, A.\ 2013, \aap, 557, A120


\bibitem[Kainulainen et al.(2011)]{kai11} Kainulainen, J., Beuther, H., Banerjee, R., et al.\ 2011, \aap, 530, A64

\bibitem[Kainulainen et al.(2009)]{kai09} Kainulainen, J., Beuther, H., Henning, T., \& Plume, R.\ 2009, \aap, 508, L35 

\bibitem[Kandori et al.(2005)]{kan05} Kandori, R., Nakajima, Y., Tamura, M., et al.\ 2005, \aj, 130, 2166 


\bibitem[Knude \& Hog(1998)]{knu98} Knude, J., \& Hog, E.\ 1998, \aap, 338, 897 



\bibitem[Lombardi(2009)]{lom09} Lombardi, M.\ 2009, \aap, 493, 735 

\bibitem[Miettinen(2012)]{mie12} Miettinen, O.\ 2012, \aap, 540, A104 

\bibitem[Mizuno et al.(1995)]{miz95} Mizuno, A., Onishi, T., Yonekura, Y., et al.\ 1995, \apjl, 445, L161 

\bibitem[Myers(2009)]{mye09} Myers, P.~C.\ 2009, \apj, 700, 
1609 

\bibitem[Ossenkopf \& Henning(1994)]{oss94} Ossenkopf, V., \& Henning, T.\ 1994, \aap, 291, 943 

\bibitem[Ostriker(1964)]{ost64} Ostriker, J.\ 1964, \apj, 140, 1056 

\bibitem[Palmeirim et al.(2013)]{pal13} Palmeirim, P., Andr{\'e}, P., Kirk, J., et al.\ 2013, \aap, 550, A38 

\bibitem[Pineda et al.(2010)]{pin10} Pineda, J.~L., Goldsmith, P.~F., Chapman, N., et al.\ 2010, \apj, 721, 686  

\bibitem[Pon et al.(2012)]{pon12} Pon, A., Toal{\'a}, J.~A., Johnstone, D., et al.\ 2012, \apj, 756, 145 

\bibitem[Pon et al.(2011)]{pon11} Pon, A., Johnstone, D., \& Heitsch, F.\ 2011, \apj, 740, 88 


\bibitem[Ragan et al.(2014)]{rag14} Ragan, S.~E., Henning, T., Tackenberg, J., et al.\ 2014, \aap, 568, A73 

\bibitem[Recchi et al.(2014)]{rec14} Recchi, S., Hacar, A., \& Palestini, A.\ 2014, \mnras, 444, 1775 

\bibitem[Recchi et al.(2013)]{rec13} Recchi, S., Hacar, A., \& Palestini, A.\ 2013, \aap, 558, A27 

\bibitem[Savage et al.(1977)]{sav77} Savage, B.~D., Bohlin, R.~C., Drake, J.~F., \& Budich, W.\ 1977, \apj, 216, 291 

\bibitem[Schmalzl et al.(2010)]{schm10} Schmalzl, M., Kainulainen, J., Quanz, S.~P., et al.\ 2010, \apj, 725, 1327

\bibitem[Schuller(2012)]{SCH12} Schuller, F.\ 2012, \procspie, 8452, 84521T 

\bibitem[Schneider \& Elmegreen(1979)]{sch79} Schneider, S., \& Elmegreen, B.~G.\ 1979, \apjs, 41, 87 

\bibitem[Schneider et al.(2010)]{sch10} Schneider, N., Csengeri, T., Bontemps, S., et al.\ 2010, \aap, 520, A49 

\bibitem[Siringo et al.(2009)]{SIR09} Siringo, G., Kreysa, E., Kov{\'a}cs, A., et al.\ 2009, \aap, 497, 945 

\bibitem[Skrutskie et al.(2006)]{skr06} Skrutskie, M.~F., Cutri, R.~M., Stiening, R., et al.\ 2006, \aj, 131, 1163    

\bibitem[Smith et al.(2014)]{smi14} Smith, R.~J., Glover, S.~C.~O., \& Klessen, R.~S.\ 2014, \mnras, 445, 2900 

\bibitem[Takahashi et al.(2013)]{tak13} Takahashi, S., Ho, P.~T.~P., Teixeira, P.~S., Zapata, L.~A., \& Su, Y.-N.\ 2013, \apj, 763, 57 

\bibitem[Vandame(2002)]{2002SPIE.4847..123V} Vandame, B.\ 2002, \procspie, 4847, 123 

\bibitem[Vilas-Boas et al.(1994)]{vil94} Vilas-Boas, J.~W.~S., Myers, P.~C., \& Fuller, G.~A.\ 1994, \apj, 433, 96 

\bibitem[Vuong et al.(2003)]{vuo03} Vuong, M.~H., Montmerle, T., Grosso, N., et al.\ 2003, \aap, 408, 581 

\end{thebibliography}
\end{document}